\documentclass[12pt]{iopart}
\pdfoutput=1 
\usepackage[utf8x]{inputenc}
\usepackage[UKenglish]{babel}
\expandafter\let\csname equation*\endcsname\relax
\expandafter\let\csname endequation*\endcsname\relax
\usepackage{amsmath}
\usepackage{verbatim}
\usepackage{iopams}
\usepackage{amssymb}
\usepackage{fullpage}
\usepackage{bm}
\usepackage{bbold}
\usepackage{appendix}
\usepackage{graphicx}
\usepackage{color}
\usepackage{tikz}
\usetikzlibrary{calc,decorations.markings}

\newcommand{\hc}{B}
\newcommand{\be}{\begin{equation}}
\newcommand{\ee}{\end{equation}}
\newcommand{\ac}{\mathcal{C}}

\newcommand{\ar}{\mathcal{R}}
\newcommand{\at}{\mathcal{T}}
\newcommand{\abb}{\mathcal{B} }
\newcommand{\abs}{\mathcal{B}^{\rm ss}}

\newcommand{\omr}{\Omega}
\usepackage{cite}
\numberwithin{equation}{section}
\allowdisplaybreaks[1]

\begin{document}
\title[]{Critical scaling in hidden state inference for linear Langevin dynamics}
\author{B Bravi$^1$\footnote{Current affiliation: Institute of Theoretical Physics, Ecole Polytechnique F\'ed\'erale de Lausanne (EPFL), CH-1015 Lausanne, Switzerland} and P Sollich$^1$}
\address{$^1$ Department of Mathematics, King's College London, Strand, London, WC2R 2LS UK}
\ead{barbara.bravi@epfl.ch and peter.sollich@kcl.ac.uk}
\date{}

\begin{abstract}
We consider the problem of inferring the dynamics of unknown (i.e.\ hidden) nodes
from a set of observed trajectories and study analytically the average prediction error 
and the typical relaxation time of correlations between errors.
We focus on a stochastic linear dynamics of continuous degrees of freedom interacting via random Gaussian 
couplings in the infinite network size limit. The expected error on the hidden time courses can be found
as the equal-time hidden-to-hidden covariance of the probability distribution conditioned on observations.
 In the stationary regime, we analyze the phase diagram in the space of relevant parameters, namely the ratio between the numbers of 
 observed and hidden nodes, the degree of symmetry of the interactions and the amplitudes of the hidden-to-hidden and hidden-to-observed couplings relative to the decay constant
 of the internal hidden dynamics. In particular, we identify critical regions in parameter space where the relaxation time and the inference error diverge, 
 and determine the corresponding scaling behaviour.
\end{abstract}

\noindent{\it Keywords: Plefka Expansion, Inference, Mean Field, Critical Scaling, Biochemical Networks, Dynamical Functional\/}\\

\maketitle
\section{Introduction}
The reconstruction of the time evolution of a system starting from
macroscopic measurements of its dynamics is a challenge of primary interest in statistical physics; see e.g.\ 
\cite{romanobattistin,inference1,inference2,roudi1}. 
The problem can be cast as follows: given the set of interaction parameters and a temporal sequence of observed variables, 
the aim is to infer the states of the variables that are unobserved or, in the terminology of machine learning, ``hidden''.

Recently, in \cite{plefkaobs}, we have proposed a method
to solve this problem for continuous degrees of freedom based on a dynamical mean-field theory, the Extended Plefka Expansion \cite{bravi}.
We specialized to the case of a linear stochastic dynamics for the purpose of a 
direct comparison with the exact computation implemented via the Kalman filter \cite{kalmanor}, a well-known
inference technique in linear (Gaussian) state space models \cite{bishop}.
In these models, the posterior distribution over the hidden dynamics is Gaussian: while the posterior mean provides the best 
estimate of the hidden dynamics, the posterior equal-time variance measures the uncertainty of this prediction and thus the inference error.
With mean-field couplings (weak and long-ranged) drawn at random from a Gaussian distribution, 
we investigated numerically the performance of the extended Plefka expansion, finding a clear improvement for large system size.
In this paper, we calculate analytically the posterior statistics in the thermodynamic limit of an infinitely large hidden system at stationarity,
with a focus on the inference error and the relaxation time, defined as
the typical timescale over which correlations between inference errors decay.
The thermodynamic limit expressions we find are expected to be exact
as we show by comparison to other methods, appealing to Random Matrix Theory and 
dynamical functionals \cite{bravikalman}. As the resulting scenario is analytically tractable, one can
study the dependence of relaxation times and the average prediction error  on key system parameters and thus 
shed light on the accuracy of the inference process. This, beyond the derivation of the general thermodynamic limit expressions, is the second major aim of this 
paper and can be seen as analogous to what has been done for learning from static data in a linear perceptron \cite{hertz,oppersolvable}, where
solvable models allowed the authors to study how the prediction error scales with the 
number of training examples and spatial dimensions.
The emphasis here is on understanding the prediction accuracy 
for hidden states from a theoretical point of view, including its dependence on macroscopic parameters
that could be measurable, at least indirectly.

The paper is organized as follows. 
In section \ref{sec:pwoEPEHN} we introduce the basic set up, a linear stochastic dynamics with hidden nodes, and we 
recall the main results of the Extended Plefka Expansion applied to this case \cite{plefkaobs}, 
as the starting point for the analysis of this paper.
In section \ref{sec:pwoTL} we consider the thermodynamic limit of infinite network size, shifting from local correlations to their macroscopic 
average across the network. The exactness of the extended Plefka predictions is shown in section \ref{sec:pwoPC} by comparison with expressions obtained 
elsewhere\cite{bravikalman} by Kalman filter and Random Matrix Theory (RMT) methods. 
From section \ref{power_spectrum}, we present a systematic analysis of these mean-field results in terms of the system properties and the number of observations.
In section \ref{sec:pwoDS} we determine the relevant dimensionless parameters governing prediction accuracy, 
namely the ratio between the numbers of observed and hidden nodes, the degree of symmetry of the hidden interactions, and the 
amplitudes of the hidden-to-hidden and hidden-to-observed interactions relative to the decay constant of the internal hidden dynamics. 
We identify critical points in this parameter space by studying the behaviour of power spectra using  a scaling approach (sections \ref{sec:pwoR}, \ref{sec:psmc} and \ref{appendix:a}). 
In the main part of the analysis we then consider the temporal correlations in the posterior dynamics of the hidden nodes,
revealing interesting long-time behaviour in the critical regions (sections \ref{sec:pwotime1}-\ref{sec:equalcorr2} and \ref{appendix:b}).
More specifically, we first present the power laws governing relaxation times in sections \ref{sec:pwotime1} and \ref{sec:pwotime2},
then discuss the corresponding correlation functions in sections \ref{sec:FT1} and \ref{sec:FT2}, and finally extract predictions
on the inference error in sections \ref{sec:equalcorr} and \ref{sec:equalcorr2}.

\section{Extended Plefka Expansion with hidden nodes}
\subsection{Set up}
\label{sec:pwoEPEHN}
Let us consider a generic network where only the dynamics of a subnetwork of nodes is observed while the others are hidden and form what we call
the ``bulk''. The indices $i,j=1,...,N^{\rm b}$ and the superscript b are used for the hidden or bulk variables; similarly the indices
$a,b=1,...,N^{\rm s}$ and the superscript s for the observed or 
subnetwork nodes of the network. Assuming linear couplings $\lbrace J_{ij} \rbrace$, $\lbrace K_{ia} \rbrace$ between hidden and observed variables $x_i$, $x_a$, 
their dynamical evolution 
is described by
\begin{subequations} 
\label{eq:lineqP}
\begin{align}
 \partial_t x_i(t)&=-\lambda x_i(t) +\sum_j J_{ij} x_j(t)+\sum_a K_{ia} x_a(t) +\xi_i(t) \label{eq:lineqP1}\\
 \partial_t x_a(t)&=-\lambda x_a(t) +\sum_b J_{ab} x_b(t)+\sum_j K_{aj} x_j(t) +\xi_a(t) \label{eq:lineqP2}
\end{align}
\end{subequations}
$J_{ij}$ (respectively $J_{ab}$) gives the hidden-to-hidden (respectively observed-to-observed) interactions, 
while the coupling between observed and hidden variables is contained in $K_{ia}$ and $K_{aj}$.
$\lambda$ is a self-interaction term acting as a decay constant and providing the basic timescale of the dynamics. 
The dynamical noises $\xi_i$, $\xi_a$ are Gaussian white noises with zero mean and diagonal covariances $\Sigma_{i}$, $\Sigma_{a}$
\begin{equation}
\langle \xi_i(t) \xi_j(t')   \rangle= \Sigma_{i}\delta_{ij} \delta(t-t')\qquad \langle \xi_a(t) \xi_b(t')   \rangle= \Sigma_{a}\delta_{ab}\delta(t-t')
\end{equation}
The application of the extended Plefka expansion to this problem with hidden and observed nodes \cite{plefkaobs}
allows us to obtain a closed system of integral equations describing the 
second moments of the Gaussian posterior distribution over the hidden dynamics, i.e.\ conditioned on the observed trajectories. These second moments are denoted by
$C_i(t,t')$, the hidden posterior variance, $R_i(t,t')$, the hidden posterior response, $B_i(t,t')$ and $B_a(t,t')$, the posterior correlations of auxiliary 
variables introduced to represent the hidden and observed dynamics respectively \emph{\`a la} 
Martin--Siggia--Rose--Janssen--De Dominicis \cite{martin,janssen,dedominicis}. They are diagonal in the site index but functions of two times by 
construction of the Extended Plefka Expansion \cite{bravi},
which gives an effectively non-interacting approximation of the dynamics where couplings among trajectories are replaced by 
a memory term (related to $R_i(t,t')$) and a time-correlated noise (whose covariance is related to $B_i(t,t')$). 

Our main interest is in the posterior statistics: the equal-time variance of inference errors gives a mean-square 
prediction error, while the time correlations between the inference errors define a posterior relaxation time. To simplify the equations derived in \cite{plefkaobs}, we consider long times, where a stationary regime is reached: all two-time functions become 
time translation invariant (TTI) and the Laplace-transformed equations yield a system of four coupled equations 
for $\tilde{C}_i(z)$, $\tilde{R}_i(z)$, $\tilde{\hc}_{i}(z)$, $\tilde{\hc}_a(z)$.
Let us briefly recall it here, as it will be the starting point for the analysis
\begin{subequations}
\label{eqs0}
\begin{align}
&-\tilde{C}_i(z)\bigg(\sum_{a}K_{ai}^2 \tilde{\hc}_{a}(z)+\sum_{j}J_{ji}^2 \tilde{\hc}_{j}(z)\bigg)+\tilde{R}_i(z)\bigg(z+\lambda-\sum_jJ_{ij}J_{ji}\tilde{R}_j(z)\bigg)=1\label{uffa1}\\
&\tilde{C}_i(z)\bigg(-z+\lambda-\sum_jJ_{ij}J_{ji}\tilde{R}_j(-z)\bigg)-\tilde{R}_i(z)\bigg(\sum_{j}J_{ij}^2\tilde{C}_j(z)+\Sigma_i\bigg)=0\label{uffa2}\\
&\tilde{\hc}_{i}(z)\bigg[\bigg(-z+\lambda-\sum_jJ_{ij}J_{ji}\tilde{R}_j(-z)\bigg)\bigg(\sum_{a}K_{ai}^2 \tilde{\hc}_{a}(z)+\sum_{j}J_{ji}^2\tilde{\hc}_{j}(z)\bigg)^{-1}\label{uffa400}\notag\\
&\qquad\quad\bigg(z+\lambda-\sum_jJ_{ij}J_{ji}\tilde{R}_j(z)\bigg)- \bigg(\sum_{j}J_{ij}^2\tilde{C}_j(z)+\Sigma_i\bigg)\bigg]=1\\
&\tilde{\hc}_a(z)\bigg[\Sigma_a+\sum_{j}K_{aj}^2\tilde{C}_j(z)\bigg]=-1\label{uffa5}
\end{align}
\end{subequations}

\subsection{Thermodynamic Limit}
\label{sec:pwoTL}
We expect the extended Plefka approach to give exact values for posterior means and variances in the case of mean-field type couplings (i.e.\ weak and long-ranged), 
in the thermodynamic limit of an infinitely large system.
More precisely we define the thermodynamic limit as the one of an infinitely large bulk and subnetwork, $N^{\rm b}, N^{\rm s} \rightarrow \infty$ at 
constant ratio $\alpha= N^{\rm s}/N^{\rm b}$. For the mean-field couplings we assume, as in \cite{bravi}, that $\lbrace J_{ij} \rbrace$ is a real matrix belonging 
to the Girko ensemble \cite{girko1}, i.e.\ its elements are independently and randomly distributed Gaussian variables with zero mean and variance satisfying
\begin{equation}
\label{jij}
 \langle {J}_{ij} {J}_{ij} \rangle=\frac{j^2}{N^{\rm b}}
\end{equation}
\begin{equation}
\label{jji}
 \langle {J}_{ji} {J}_{ij} \rangle=\frac{\eta\, j^2}{N^{\rm b}}
\end{equation}
The parameter $\eta\in[-1,1]$ controls the degree to which the matrix $\lbrace J_{ij} \rbrace$ is symmetric. In particular, the dynamics is 
non-equilibrium -- it does not satisfy detailed balance -- whenever $\eta < 1$.
Similarly, $\lbrace K_{ai} \rbrace$ is taken as a random matrix with uncorrelated zero mean Gaussian entries of variance
\begin{equation}
\label{kia}
\langle {K}_{ai} \rangle=0 \qquad \langle {K}_{ai}^2 \rangle=\frac{k^2}{N^{\rm b}}
\end{equation}
We have introduced amplitude parameters $j$ for $\lbrace J_{ij} \rbrace$ (hidden-to-hidden) and $k$ for $\lbrace K_{ai} \rbrace$ (hidden-to-observed) here.
The scaling of both types of interaction parameters with  $1/\sqrt{N^{\rm b}}$ ensures that, when the size of the 
hidden part of the system increases, the typical contribution it makes to the time evolution of each hidden and observed 
variable stays of the same order.
The high connectivity, where all nodes interact with all other ones, implies that in the thermodynamic limit all nodes 
become equivalent. Local response and correlation functions therefore become identical to their averages over nodes, defined as
\begin{eqnarray}
 \tilde{R}^{\rm bb|s}(z)=\frac{1}{N^{\rm b}}\sum_{j}\tilde{R}_j(z)\\
 \tilde{C}^{\rm bb|s}(z)=\frac{1}{N^{\rm b}}\sum_{j}\tilde{C}_j(z)\\
 \tilde{\hc}^{\rm bb|s}(z)=\frac{1}{N^{\rm b}}\sum_{j}\tilde{\hc}_{j}(z) 
\end{eqnarray}
As a consequence, all site indices can be dropped and the correlation and response functions can be replaced by their mean values,
$\tilde{C}_i(z)\equiv \tilde{C}^{\rm bb|s}(z)$, $\tilde{R}_i(z)\equiv \tilde{R}^{\rm bb|s}(z)$, $\tilde{\hc}_{i}(z)\equiv \tilde{\hc}^{\rm bb|s}(z)$. 
The superscripts bb$|$s here emphasize that we are considering moments conditioned on subnetwork values
though for brevity we drop them in the rest of the calculation. Similarly, for the correlations of auxiliary variables related to observations we can set
$\tilde{\hc}_{a}(z)\equiv \tilde{B}^{\rm ss}(z)$.
Of primary interest is then $\tilde{C}(z)$, the Laplace transformed posterior (co-)variance function of prediction errors, which as a site-average can be thought of 
as a macroscopic measure of prediction performance. This should become self-averaging in the thermodynamic limit. 
To see this, consider e.g.\ the sum $
\sum_{j}J_{ij}J_{ji}\tilde{R}_j(z)$. Replacing $\tilde{R}_j$ by $\tilde{R}$, the prefactor is a sum of $N^{\rm b}$ terms so converges to its average $N^{\rm b} \eta j^2/N^{\rm b} = \eta j^2$ in the thermodynamic limit. 
So, as in \cite{bravi}, we can replace
\begin{displaymath}
 \sum_{j}J_{ij}J_{ji}\tilde{R}_j(z)\sim \eta\, j^2 \tilde{R}(z)
 \end{displaymath} 
Making this and similar substitutions in the system \eqref{eqs0}, and choosing scalar noise covariances $\Sigma_i=\sigma_{\rm b}^2$ and $\Sigma_a=\sigma_{\rm s}^2$  as in \cite{bravikalman}, one finds
\begin{subequations}
\label{eq:PlefkasysTL}
\begin{align}
 &-\tilde{C}(z)\bigg(-\frac{\alpha k^2}{(\sigma_{\rm s}^2+ k^2 \tilde{C}(z))}+j^2 \tilde{\hc}(z)\bigg)+
\tilde{R}(z)\bigg(z+\lambda-\eta j^2 \tilde{R}(z)\bigg)=1\label{eq:PlefkasysTL1}\\
 &\tilde{C}(z)\bigg(-z+\lambda-\eta j^2 \tilde{R}(-z)\bigg) - \tilde{R}(z)\bigg(j^2 \tilde{C}(z)+\sigma_{\rm b}^2\bigg)=0 \label{eq:PlefkasysTL2}\\
\begin{split}
 &\tilde{\hc}(z)\bigg[\bigg(-z+\lambda-\eta j^2 \tilde{R}(-z)\bigg)\bigg(-\frac{\alpha k^2}{(\sigma_{\rm s}^2+ k^2 \tilde{C}(z))}+j^2 \tilde{\hc}(z)\bigg)^{-1}
 \bigg(z+\lambda-\eta j^2 \tilde{R}(z)\bigg) \label{eq:PlefkasysTL3}\\
&\qquad -\bigg(j^2 \tilde{C}(z)+\sigma_{\rm b}^2\bigg)\bigg]=1
\end{split}
\end{align}
\end{subequations}
Here $\tilde{\hc}^{\rm ss}(z)=-1/(\sigma_{\rm s}^2+ k^2 \tilde{C}(z))$ has already been substituted into equations \eqref{eq:PlefkasysTL1} 
and \eqref{eq:PlefkasysTL3}.

We comment briefly on the relation of the above results to the Fluctuation Dissipation Theorem (FDT), which in terms of Laplace transforms reads
\begin{equation}
\label{eq:pwoFDT}
 z\tilde{C}(z)=-\frac{\sigma_{\rm b}^2}{2}\big[\tilde{R} (z)-\tilde{R} (-z) \big]
\end{equation}
This can be compared with the expression for $\tilde{R}(z)$ that follows from \eqref{eq:PlefkasysTL2} (taken for $z$ and $-z$)
\begin{equation}
 \tilde{R} (z)=\tilde{C} (z)\bigg[\frac{\lambda}{\sigma_{\rm b}^2+j^2(1+\eta)\tilde{C}(z)}-\frac{z}{j^2(1-\eta)\tilde{C}(z)+\sigma_{\rm b}^2}\bigg]
\label{eq:Rsol}
\end{equation}
The r.h.s.\ of the FDT is then
\begin{equation}
-\frac{\sigma_{\rm b}^2}{2}\big[\tilde{R} (z)-\tilde{R} (-z) \big]
=z\tilde{C}(z) \frac{\sigma_{\rm b}^2 }
{j^2(1-\eta)\tilde{C}(z)+\sigma_{\rm b}^2}
\end{equation}
Comparing with \eqref{eq:pwoFDT}, the FDT is satisfied for symmetric couplings ($\eta=1$) as expected, while there 
are progressively stronger deviations from FDT as $\eta$ decreases towards $-1$.

\subsection{Comparison with known results}
\label{sec:pwoPC}
As a consistency check and to support our claim of exactness in the thermodynamic limit, we briefly compare our results with expressions for 
$\tilde{C} (z)$ and $\tilde{R} (z)$ that can be worked out by alternative means.

In general, from \eqref{eq:PlefkasysTL2} we can get an expression for $\tilde{R} (z)$ in terms of $\tilde{C} (z)$: this is \eqref{eq:Rsol}.
Substituting into \eqref{eq:PlefkasysTL1}, $\tilde{\hc} (z)$ can also be worked out as a function of $\tilde{C} (z)$. Using these expressions for $\tilde{R} (z)$ 
and $\tilde{\hc} (z)$ in equation \eqref{eq:PlefkasysTL3}, one finds a closed algebraic equation for the posterior variance $\tilde{C}(z)$
\be
\label{algcorr_gen}
z^2= \left[-\frac{\sigma_{\rm b}^2}{\tilde{C}} +\frac{ \alpha \frac{k^2 \sigma_{\rm b}^2}{\sigma_{\rm s}^2}}{1+ \frac{k^2}{\sigma_{\rm s}^2}\tilde{C}}
+\frac{j^2}{1+\frac{j^2}{\sigma_{\rm b}^2}\tilde{C}}+ 
\frac{\lambda^2}{\bigg(1+(1+\eta)\frac{j^2}{\sigma_{\rm b}^2}\tilde{C}\bigg)^2}\right]
\bigg(1+(1-\eta)\frac{j^2}{\sigma_{\rm b}^2}\tilde{C}\bigg)^2
\ee
This is the same expression as obtained by calculations in \cite{bravikalman} using an explicit average over the quenched disorder variables $J_{ij}$ and $K_{ia}$.
Particular cases are also further validated by random matrix theory, as follows.

\subsubsection{$\alpha=0$.}
\label{sec:a0}
This case corresponds to the absence of observations. One has then $\tilde{\hc}_{i}(z)$, $\tilde{\hc}_{a}(z)\equiv 0$ as these quantities simply play the role of Lagrange multipliers 
enforcing the conditioning on observations. 
To see this formally from the $\alpha \to 0$ limit of \eqref{eq:PlefkasysTL} one sets $\tilde{\hc}_{i}(z)=\alpha \tilde{D}_{i}(z)$ and
$\tilde{\hc}_{a}(z)=\alpha\tilde{D}_{a}(z)$ where the $\tilde{D}$ stay nonzero for $N^{\rm s}\to 0$.
One verifies that under this assumption the system has as solution the responses and correlations known from \cite{bravi} (where a thorough analysis of the thermodynamic 
limit of an analogous linear dynamics \emph{without} observations was provided)
\begin{equation}
 \label{eq:PleRes}
 \tilde{R}(z)= \frac{1}{2\eta}(z+\lambda)-\frac{1}{2\eta}\sqrt{(z+\lambda)^2-4j^2\eta}
\end{equation}
\begin{equation}
 \label{FinCor}
 \tilde{C}(z)=\frac{4\,\sigma_{\rm b}^2}{\big[(\lambda+z)+\sqrt{(\lambda+z)^2-4j^2\eta}\big]\big[(\lambda-z)+\sqrt{(\lambda-z)^2-4j^2\eta}\big]-4 j^2}
\end{equation}

\subsubsection{$j=0$.}
\label{sec:j0}
In this situation, the hidden variables have got no direct interactions. By solving \eqref{algcorr_gen}, one obtains
\small
\begin{equation}
\label{eq:osi}
\tilde{C} (z)=\frac{\sigma_{\rm s}^2}{2 k^2}\frac{\sigma^2}{{(\lambda^2-z^2)}}\bigg\lbrace 1-\alpha-
\bigg(\frac{\lambda^2-z^2}{\sigma^2}\bigg)+\sqrt{\bigg[1-\alpha- \bigg(\frac{\lambda^2-z^2}{\sigma^2}\bigg)\bigg]^2+
4\bigg(\frac{\lambda^2-z^2}{\sigma^2}\bigg) }\bigg\rbrace
\end{equation}
\normalsize
where $\sigma=\sigma_{\rm b} k/\sigma_{\rm s}$. 
This coincides with the result in \cite{bravikalman}, which can be derived separately using methods from random matrix theory.

\subsubsection{$\eta=1$.}
\label{sec:eta1}
Here we have symmetric hidden-to-hidden couplings and \eqref{algcorr_gen}
can be simplified to
\begin{equation}
\label{eq:Plefkasymm}
  z^2=-\frac{\sigma_{\rm b}^2}{\tilde{C} (z)}+\frac{\alpha\frac{k^2 \sigma_{\rm b}^2}{\sigma_{\rm s}^2}}{1+\frac{k^2}{\sigma_{\rm s}^2}\tilde{C} (z)}+
  \frac{j^2}{1+\frac{j^2}{\sigma_{\rm b}^2}\tilde{C} (z)}+\frac{\lambda^2}{\bigg(1+2\frac{j^2}{\sigma_{\rm b}^2}\tilde{C} (z)\bigg)^2}
\end{equation}
This fifth order equation for the single site posterior covariance predicted
by the extended Plefka expansion with hidden nodes is again confirmed by random matrix theory results \cite{bravikalman}.

\section{Critical regions}
\label{power_spectrum}
We can now proceed to the main contribution of this work, namely the study of the properties of our conditioned dynamical system. The focus
will be first on the power spectrum of the posterior covariance, given by $\tilde{C} (\text{i}\omega)$, the Laplace 
transform $\tilde{C}(z)$ evaluated at $z=\text{i}\omega$, as this is the quantity that is immediately computable from the equations \eqref{eq:PlefkasysTL}. 
We will then translate the insights from the frequency domain analysis into the time domain, to access key observables including the timescales of posterior correlations.

We use the formulae provided by the Plefka approach for our analysis but stress that the behaviour we find, including the presence of critical regions 
in parameter space, is not an artefact of the Plefka expansion. The Plefka expansion
is known to have a finite radius of convergence \cite{plefka} in terms of coupling strength, and so it might be  thought that one should check whether our parameter ranges lie in the 
region of validity of the expansion. However, as highlighted in section \ref{sec:pwoPC}, the Plefka results \eqref{eq:PlefkasysTL} fully agree with other methods that do \emph{not} rely on perturbative expansions, namely random matrix theory and
disorder-averaged dynamical functionals \cite{bravikalman}, which implies that there can be no issues with convergence of the Plefka expansion.

\subsection{Dimensionless system for the power spectrum}
\label{sec:pwoDS}

We would like first to understand how $\tilde{C}(\text{i}\omega)$ depends on the parameters $\lambda$, $j$, $k$, $\sigma_{\rm s}$, $\sigma_{\rm b}$, $\eta$ and $\alpha$. 
The last two of these are already dimensionless. By extracting the appropriate dimensional scales, we can reduce the other five parameters to only two 
dimensionless combinations.

From \eqref{eq:lineqP} one sees that $j$, $k$, $\lambda$ have dimensions $t^{-1}$, while the dimension of $\sigma_{\rm s/b}^2$ is $x^2 t^{-1}$. We can build from these the dimensionless parameters $\gamma=j/\lambda$ and $p=\lambda/\sigma$. Here $\sigma=\sigma_{\rm b} k/\sigma_{\rm s}$, which has dimension $t^{-1}$ and contains the observation ``intensity'' $k$ as well as the ratio between the dynamical noises $\sigma_{\rm b}/\sigma_{\rm s}$. The latter is a third dimensionless parameter but as it only enters one prefactor we will not need to keep it separately.

Extracting appropriate dimensional amplitudes for all four two-point functions, we write them as 
 \begin{subequations}
 \label{eq:adimmom}
 \begin{align}
  \tilde{C} (\text{i}\omega)=&\frac{\sigma_{\rm s}^2}{k^2}\ac_{\alpha,\gamma,\eta,p}(\omr)\label{eq:adim1}\\
  \tilde{R} (\text{i}\omega)=& \frac{1}{j} \ar_{\alpha,\gamma,\eta,p}(\omr)\\
  \tilde{B} (\text{i}\omega)=&\frac{1}{\sigma_{\rm b}^2}\abb_{\alpha,\gamma,\eta,p}(\omr)\\
 \tilde{B}^{\rm ss}(\text{i}\omega)=&\frac{1}{\sigma_{\rm s}^2}\abs_{\alpha,\gamma,\eta,p}(\omr)
\end{align}
\end{subequations}
Here $\Omega=\omega/\sigma$ is a dimensionless frequency; similarly
$\ac_{\alpha,\gamma,\eta,p}(\omr)$, $\ar_{\alpha,\gamma,\eta,p}(\omr)$, $\abb_{\alpha,\gamma,\eta,p}(\omr)$ and 
$\abs_{\alpha,\gamma,\eta,p}(\omr)$ are dimensionless and depend on the dimensionless parameters $\alpha$, $\gamma$, $\eta$ and $p$: for the sake of brevity, we do not write the subscripts indicating this dependence in the following.
Let us briefly comment on \eqref{eq:adim1}. One sees that $\tilde{C}(z)$ is directly proportional to $\sigma_{\rm s}^2$ and inversely proportional to $ k^2$: the weaker the hidden-to-observed coupling and the 
stronger the dynamical noise acting on the observed variables, the less information one can extract from the subnetwork trajectories and the more uncertain the predictions for the behaviour of the bulk.

To summarize, we switch from eight original parameters $\lbrace \alpha, \eta, \lambda, j, k, \sigma_{\rm s}, \sigma_{\rm b},\omega\rbrace$ to a set of five dimensionless parameters
$\lbrace \alpha, \eta, \gamma, p, \omr\rbrace$. For the dimensionless second moments \eqref{eq:adimmom}, the system \eqref{eq:PlefkasysTL} becomes  
\begin{subequations}
\label{eq:adimplefka}
\begin{align}
 &-\ac\bigg(-\frac{\alpha}{1+ \ac}+(\gamma p)^2\abb\bigg)+(\gamma p)^{-1}\ar\bigg(\text{i}\omr+p-\eta\,\gamma\, p\,\ar\bigg)=1\label{eq:adimplefka1}\\
 &(\gamma p)\,\ac\bigg(-\text{i}\omr+p-\eta\, \gamma\, p\, \ar_{-}\bigg)-\ar\bigg((\gamma p)^2\, \ac+1\bigg)=0\label{eq:adimplefka2}\\
 &\abb\bigg[\bigg(-\text{i}\omr+p-\eta\,\gamma\, p\,\ar_{-}\bigg) \bigg(-\frac{\alpha}{1+ \ac}+(\gamma p)^2\abb\bigg)^{-1}\bigg(\text{i}\omr+p-\eta\,\gamma\, p\,\ar\bigg)
-(\gamma p)^2 \ac-1\bigg]=1\label{eq:adimplefka3}
\end{align}
\end{subequations}
where we have dropped the frequency argument and introduced the shorthand $\ar_{-}=\ar(-\omr)$.
The solution for $\abs$, already taken into account by substitution in \eqref{eq:adimplefka}, is given by $-1/(1+\ac)$.

\subsection{Critical scaling}
\label{sec:pwoR}
\begin{figure}
\includegraphics[width=0.9\textwidth]{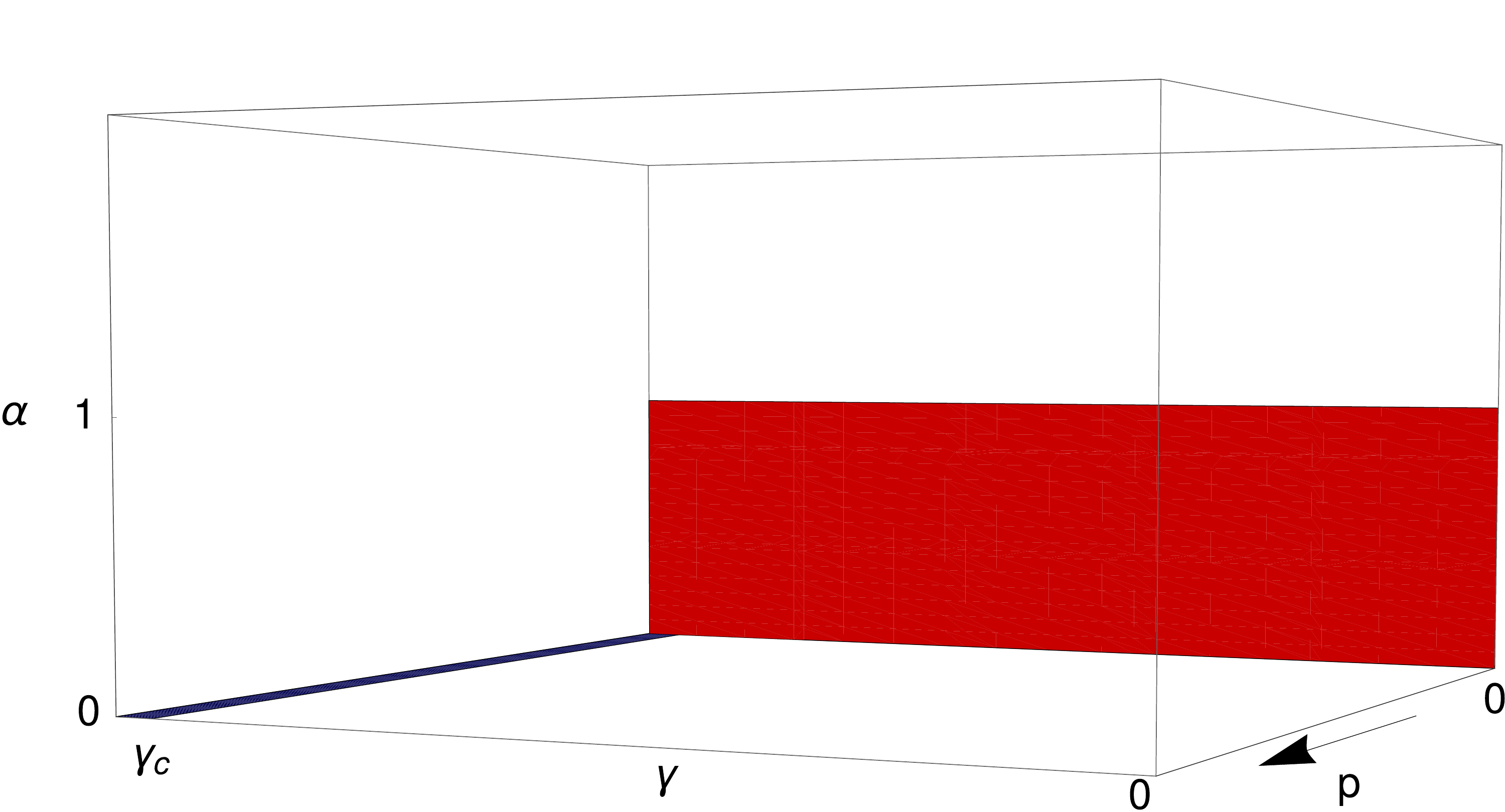}
\caption{Parameter space spanned by $\alpha$, $\gamma$, $p$. The blue stripe and the red area mark the values for which the posterior covariance $\ac(0)$ becomes singular, 
i.e.\ respectively $\alpha=0$, $\gamma>\gamma_c$ ($\forall p$) and $p=0$, $0<\alpha<1$ ($\forall \gamma$).} 
\label{fig:gammapalpha}
\end{figure}
We next analyze in the parameter space $\alpha, \gamma, p$ (at fixed $\eta$) the singularities of $\ac(0)$, the (dimensionless) zero frequency posterior covariance. 
$\ac(0)$ is a convenient quantity to look at in the first instance as it helps one detect where qualitative changes of behaviour occur. We will see that these then show up also in
the relaxation time and the inference error itself, i.e.\ the equal time posterior variance $C(t-t)=C(0)$.
Their behaviour will be addressed respectively in sections \ref{sec:pwotime1}, \ref{sec:pwotime2} and \ref{sec:equalcorr}, \ref{sec:equalcorr2},
while in sections \ref{sec:FT1} and \ref{sec:FT2} we clarify further the interpretation of our results for the relaxation times by looking at the shape of the correlation functions.

Independently of $\eta$, we find two critical regions that are shown graphically in figure \ref{fig:gammapalpha}:
\begin{enumerate}
\item $\forall p$, $\alpha=0$ and $\gamma>\gamma_c$
\item $\forall \gamma$, $p=0$ and $0<\alpha<1$ 
\end{enumerate}
The first case gives back the dynamics without observations ($\alpha=0$), for which $\gamma<\gamma_c=1/(1+\eta)$ is the condition of stability beyond 
which trajectories typically diverge in time (see \cite{bravi}). 
Interestingly, as soon as $\alpha>0$, the constraints from observations make the solution stable irrespective of whether $\gamma$ is smaller or bigger than 
the critical value.
For $\gamma>\gamma_c$ the observed trajectories would then be divergent, and so would the predicted hidden trajectories, while the error (posterior variance) of the predictions would remain bounded. It is difficult to conceive of situations where divergent mean trajectories would make sense, however, so we 
only consider the range $\gamma\leq\gamma_c$ in our analysis. In what follows, we will plot all the curves
that refer to this region of the parameter space in blue.

The second limit, $p\to 0$, corresponds, for fixed ratio between noises $\sigma_{\rm s}/\sigma_{\rm b}$, to $k\gg\lambda$: 
we call this scenario an ``underconstrained'' hidden system. In general for large $k$ the posterior variance decreases as $1/k^2$,
as used in the scaling \eqref{eq:adim1}. But for $\alpha<1$ there are directions in the space of hidden 
trajectories that are not constrained at all by subnetwork observations, and their variance will scale 
as $1/\lambda^2$ instead. These directions give a large contributions to the dimensionless $\mathcal{C}(\Omega)$ 
that diverges in the limit $k\gg \lambda$. In general one has a similar effect 
when $k/\sigma_{\rm s}\gg \lambda/\sigma_{\rm b}$, where the noise in the dynamics acts to effectively reduce the relevant interaction or decay constant.
This behaviour is broadly analogous to what happens in learning of linear functions from static data
 \cite{hertz,oppersolvable}: there the prediction error will also diverge when $\alpha$, which is 
 then defined as the ratio of number of examples to number of spatial dimensions, is less than unity and no regularization is applied. 
 For the curves belonging to this second region we choose the colour red.

Close to the two critical regions in the space of $\alpha$, $\gamma$ and $p$, $\ac(0)$ is expected to exhibit a power-law dependence on the parameters specifying the 
distance away from the critical point. 
One can study this behaviour by using standard scaling techniques developed for the study of critical phase transitions \cite{cardy}. 
The aim of this analysis, which we describe in \ref{appendix:a} and 
summarize in the next section, is to find the master curves and associated exponents that describe the approach to 
the critical point(s). From the master curves for the power spectrum we can then derive (\ref{appendix:b}) the corresponding scaling behaviour of the relaxation
times and the inference error, which is discussed in sections 
\ref{sec:pwotime1}--\ref{sec:equalcorr2} below.

\subsection{Power Spectrum Master Curves}
\label{sec:psmc}
\begin{figure}
\includegraphics[width=0.9\textwidth]{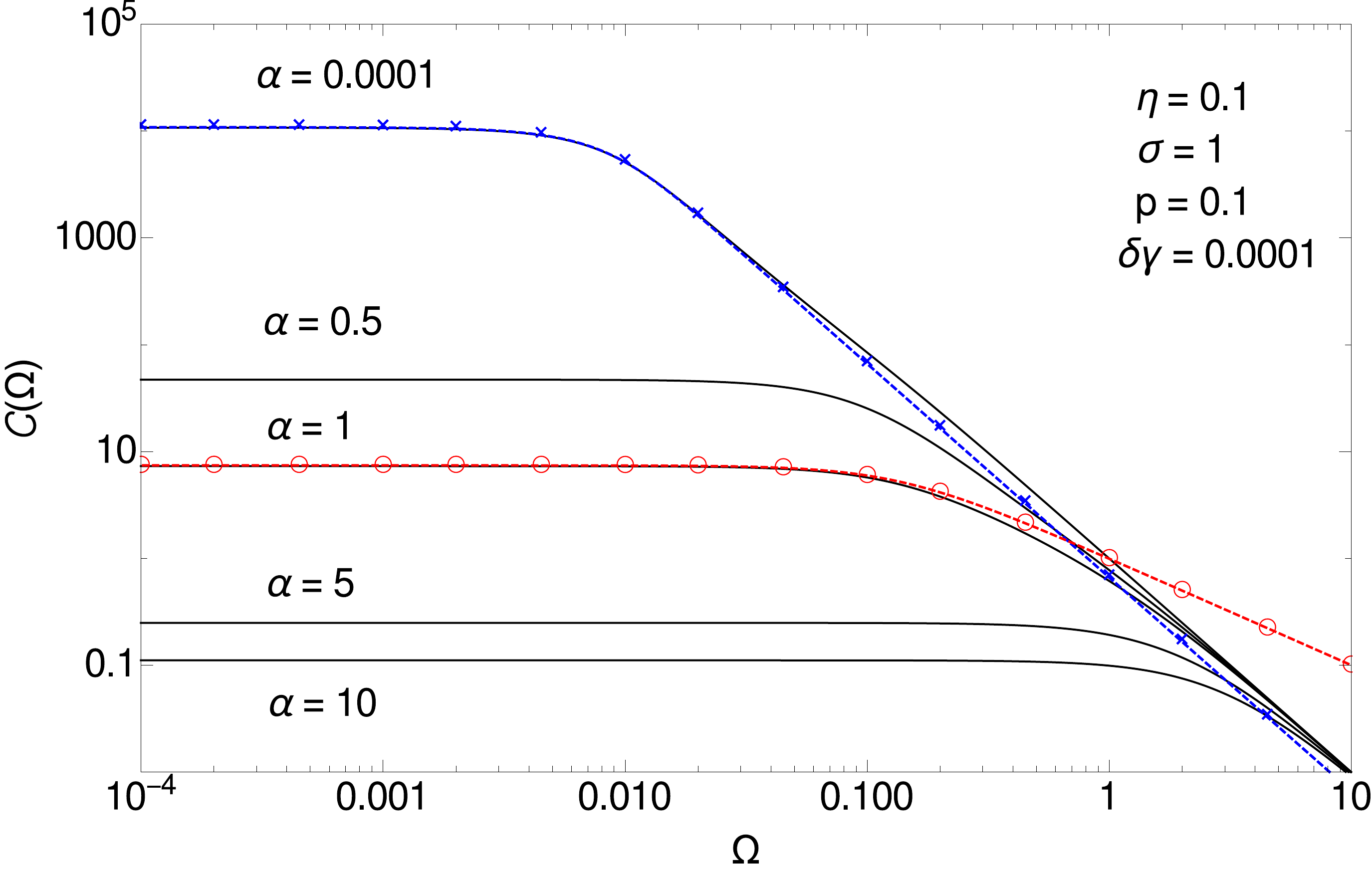}
\caption{$\ac(\omr)$ for different $\alpha$, at small fixed $\eta$. We have chosen $\gamma$ and $p$ close to their critical values $1+\eta$ and $0$, respectively, 
in order to see both critical regions as $\alpha$ is increased.  
The master curves for $\alpha\to 0$ and $\alpha\to 1$, resulting from the critical rescalings in \ref{appendix:a}, 
are plotted (blue dashed line with crosses and red dashed line with circles respectively): $\omr^{*}\sim 10^{-3}$ for the $\alpha\to 0$ master curve, 
while $\omr^{*}\sim 10^{-1}$ for the one for $\alpha\to 1$. For $\omr^{*}\ll\omr\ll 1$ one has a Lorentzian tail $\ac\sim 1/\omr^2$ for 
$\alpha$ small, while a different power-law feature, namely $\ac\sim 1/\omr$, emerges for $\alpha\to 1$. 
At $\omr\sim O(1)$ a crossover to Lorentzian behaviour is seen for any $\alpha$.} 
\label{fig:plot0}
\end{figure}
The master curves of the power spectrum 
describe its shape in the critical regions and for small frequencies $\Omega$ of the order of a parameter-dependent characteristic frequency 
$\Omega^{*}$ (see \ref{appendix:a}). These master curves are useful because they highlight the two main trends (with $\eta$ and with $\alpha$) that will
manifest themselves in 
the relaxation time and in the inference error. 
In the first critical region where there are few observations, 
the conditional dynamics is dominated by hidden-hidden interactions. 
Their degree of  symmetry $\eta$ therefore plays a key role and in fact determines a non-trivial crossover in the critical behaviour (see figure \ref{fig:plotebar}).
As developed in \ref{sec:pwoMC1}, this crossover from equilibrium to generic non-equilibrium dynamics can be studied by 
including $\epsilon = 1-\eta$ as a small parameter in the scaling analysis.
By contrast, in the second critical region 
the scaling functions do not depend on $\eta$ (see figure \ref{fig:alpha0}, right). The intuition here
is that the critical behaviour is dominated by whether a direction in the hidden trajectory space is constrained by observations or not, i.e.\ by hidden-to-observed rather 
than hidden-to-hidden interactions.

Regarding the trend with $\alpha$, the main feature is the decrease in $\ac(0)$ with increasing $\alpha$
meaning that with many observations the predictions for the hidden dynamics will become arbitrarily precise, in line with intuition. This can be seen 
in figure \ref{fig:plot0}, where one observes also that at $\omr\sim O(1)$ all spectra 
collapse into a Lorentzian tail. This indicates an exponential decay of the correlations between prediction errors in the temporal domain.
As the amplitude of the tail is largely $\alpha$-independent, the typical time of this exponential decay decreases with $\alpha$: with 
many observations, errors in the prediction of the hidden states become progressively less correlated with each other.  

This already shows that from the power spectrum $\ac(\omr)$
one can extract useful information on relaxation times and the same is true for the inference error, as 
we analyze more systematically in the next sections. 
Importantly, we find nontrivial power-law dependences on $\alpha$ that arise from the dynamical nature of 
our problem and have no analogue in static learning~ \cite{hertz,oppersolvable}.

\subsection{Relaxation time for $\gamma\to\gamma_c$ and $\alpha\to 0$}
\label{sec:pwotime1}
We look at the relaxation time, which is a measure of time correlations in the errors of inferred hidden values. 
We study in particular how it depends on the number of observations and the interaction parameters. 

The relaxation time can be defined in a mean-squared sense as 
\be
\label{eq:time}
\tau^2=\frac{\int_{-\infty}^{+\infty} t^2 C (t)dt}{2\, \tilde{C} (0)}
=-\frac{1}{2\, \ac(0)}\frac{d^2 \ac(\omega)}{d^2\omega}\bigg\lvert_{\omega=0}
\ee
Given this relation, results for power spectra can be directly used to obtain the master curves for relaxation times, see \ref{sec:pwotimea1app}
and \ref{sec:pwotime2app}. To construct these master curves,
let us introduce the dimensionless version of the relaxation timescale, 
\be
\label{eq:timeadim}
\at=\sigma\tau 
\ee
We look separately at the two critical regions, starting in this subsection with the first one, i.e.\ in the vicinity of the ``no observations'' critical point. We summarize here the main findings, leaving details for \ref{sec:pwotimea1app}. The distance from the singularity is controlled by $\alpha$ itself and $\delta\gamma=\gamma_c-\gamma$.
As mentioned in the overview in section \ref{sec:psmc}, the behaviour in this region depends crucially on the symmetry parameter $\eta$ and so we break down 
the results accordingly.
\subsubsection{$\eta=1$.}
For $\delta\gamma^2 < \alpha\ll 1$ we find
\be
\label{timealpha0mt}
\at\sim \alpha^{-\frac{2}{3}}
\ee
This power law dependence is visible in figure \ref{fig:plottot} (left, see curve for $\eta=1$).

\subsubsection{$-1<\eta<1$.}
The main result here is
\be
\label{timealpha1mt}
\at\sim \alpha^{-\frac{1}{4}} \qquad \delta\gamma^2 < \alpha \ll 1
\ee
This dependence can be verified in figure \ref{fig:plottot} (left, see curve for $\eta=0.1$).

\subsubsection{Crossover at $\eta\approx 1$.}
For symmetries $\eta$ close to 1, a crossover between the 
$\alpha^{-\frac{2}{3}}$ and $\alpha^{-\frac{1}{4}}$ behaviours occurs at a value of $\alpha^{*}$ dependent on $\epsilon = 1-\eta$.
This dependence can be analyzed quantitatively, see \ref{sec:crossover_time}; for an illustration we refer to the $\eta=0.85$ 
curve in figure \ref{fig:plottot} (left).

\subsection{Relaxation time for $\alpha \to 1$ and $p \to 0$}
\label{sec:pwotime2}
\begin{figure}
\includegraphics[width=0.485\textwidth]{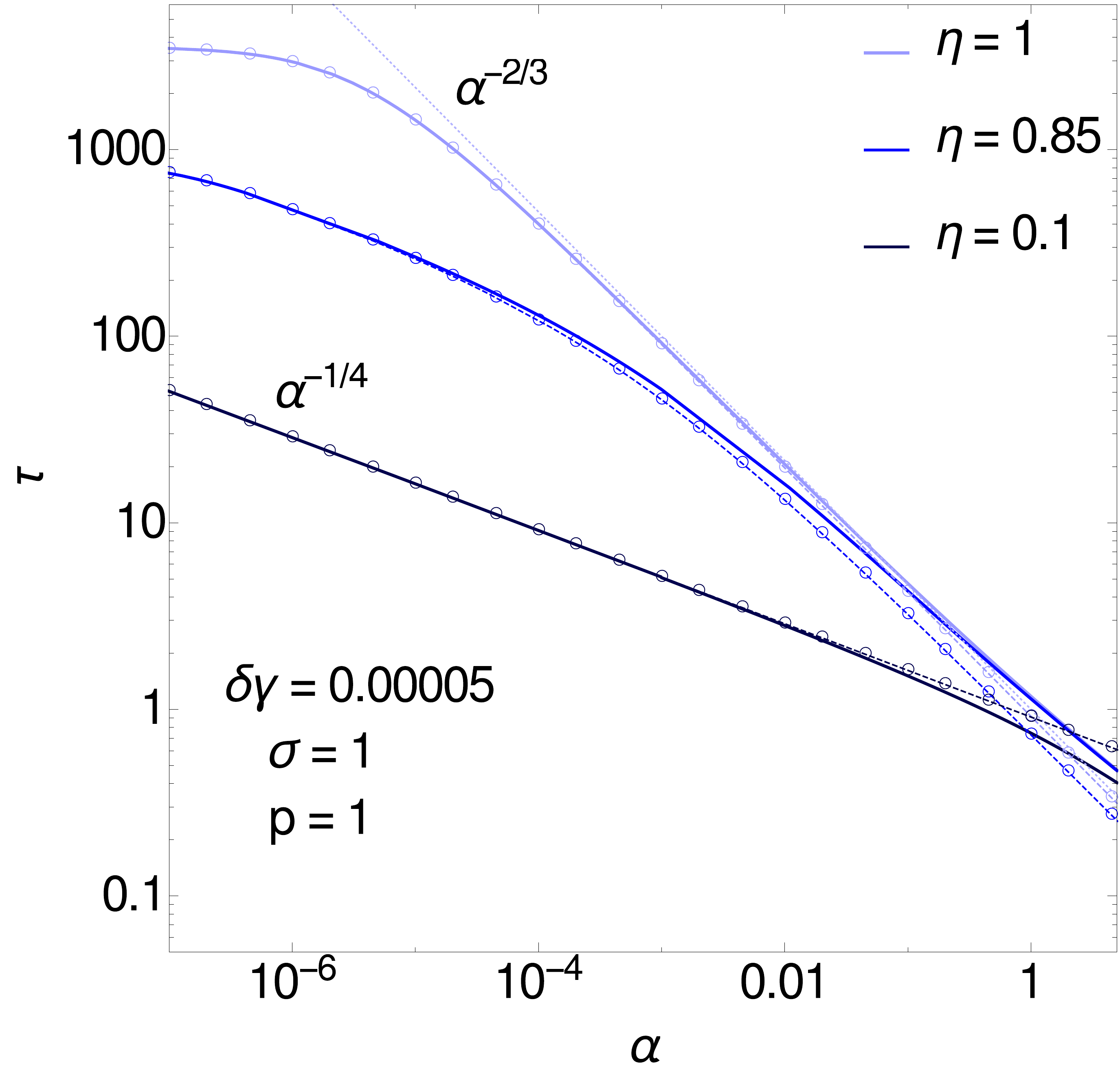} 
\includegraphics[width=0.485\textwidth]{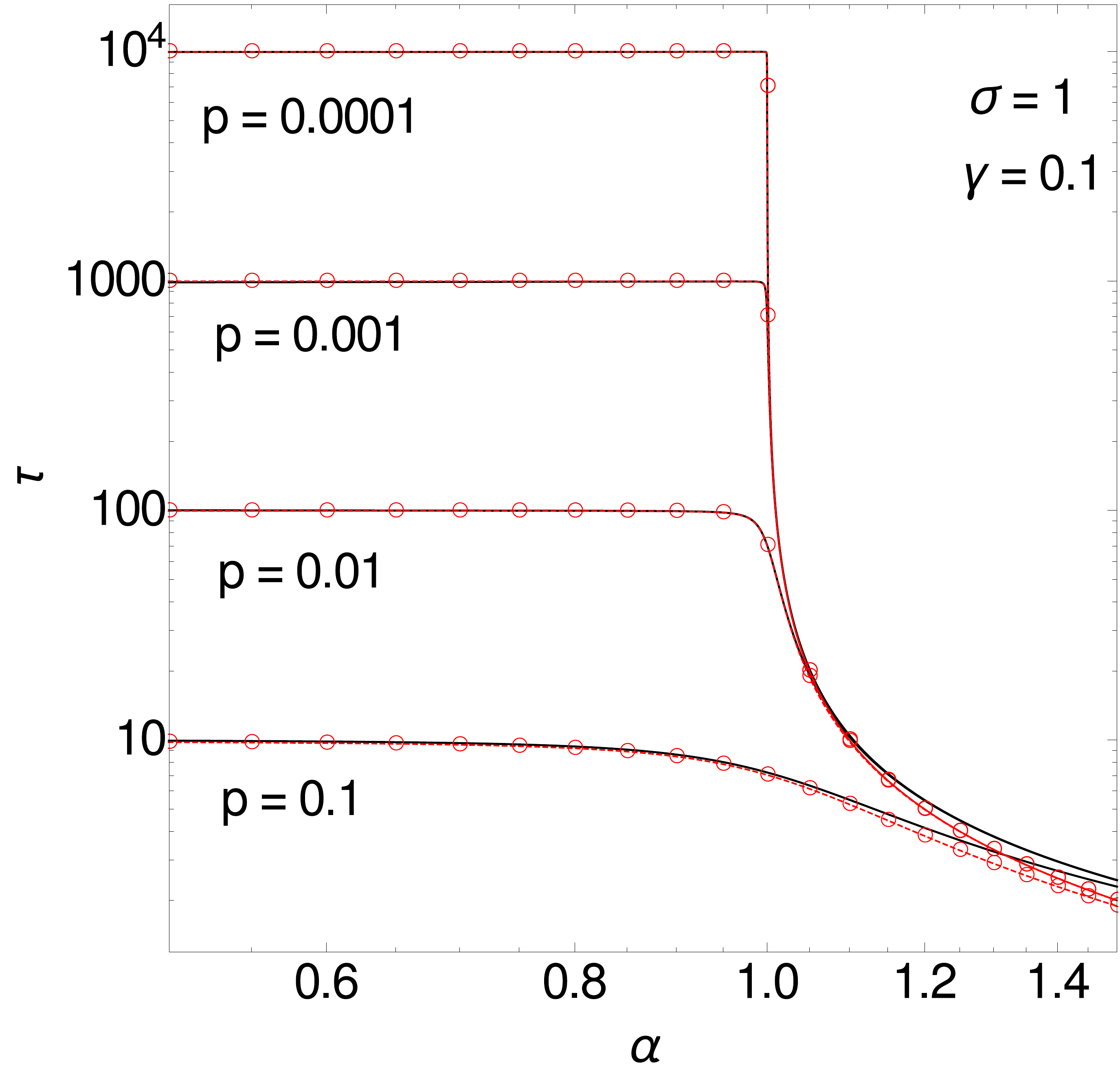}
\caption{(Left) Relaxation time in the vicinity of the critical region $\alpha \to 0$ and $\gamma \to \gamma_c$, as a function of $\alpha$ and for different $\eta$.
Solid lines are the numerics, dashed ones with circles the analytic master curves. A plateau for small $\alpha$ emerges for $\eta$ close to 1. 
For $\delta\gamma^2< \alpha\ll 1$ one can see $\tau \sim \alpha^{-\frac{2}{3}}$ for $\eta=1$ and $\tau \sim \alpha^{-\frac{1}{4}}$ for $\eta=0.1$, consistent with 
Eqs.~\eqref{timealpha0mt} and \eqref{timealpha1mt}. The case $\eta=0.85$ interpolates between these power 
tails with a crossover at $\alpha^{*} \sim 0.0005$. (Right) Relaxation time in the vicinity of the critical region $\alpha \to 1$ and $p \to 0$, as a function of $\alpha$. 
As the variation with $\gamma$ is weak in this regime, we fix $\gamma=0.1$.
Similarly the results are largely independent of $\eta$; we choose $\eta=0.1$.
Solid lines are the numerics, dashed red ones with circles the analytic master curves. One can see that $\tau$ stays roughly constant 
for $\alpha<1$ while it drops as $\approx\sigma/(\alpha-1)$ for larger $\alpha$ in line with Eq.~\eqref{eq:timep0}. }
\label{fig:plottot}
\end{figure}
\begin{figure}
\includegraphics[width=0.48\textwidth]{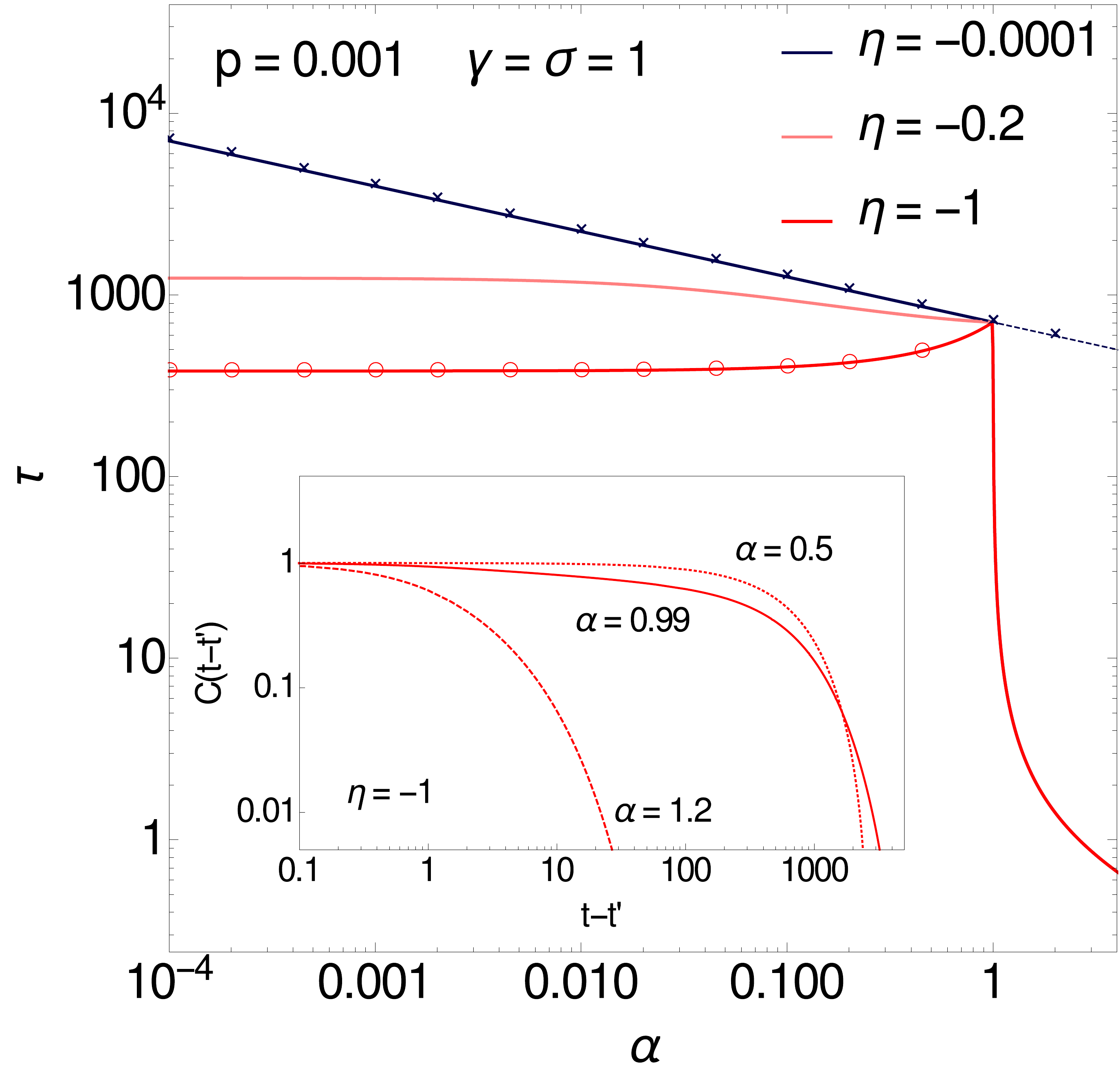}
\includegraphics[width=0.47\textwidth]{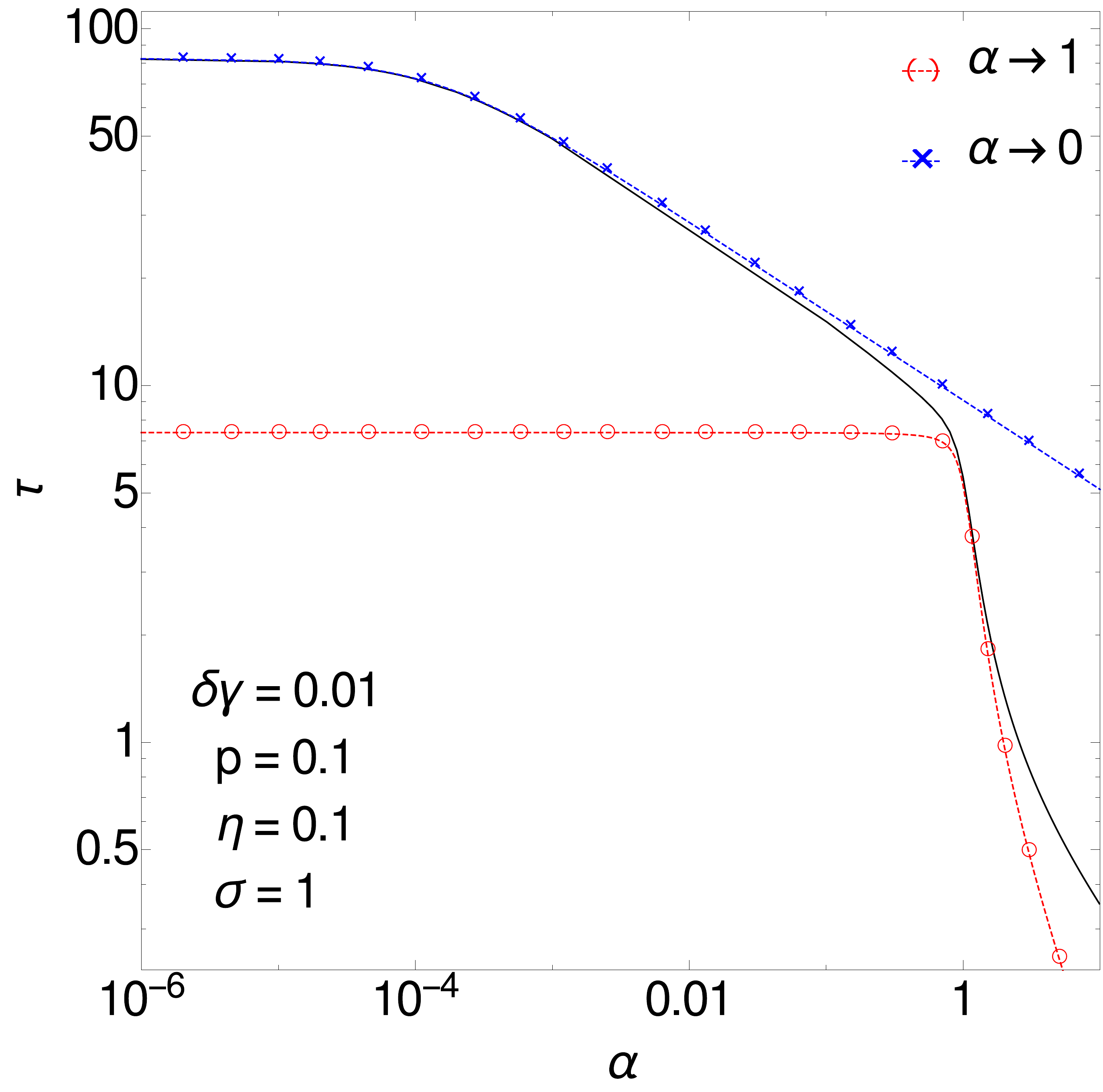}
\caption{(Left) 
Relaxation times $\tau$ as a function of $\alpha$ for small $p$: we focus on negative symmetries and we fix $\gamma=1$.
For $\eta=-1$, $\gamma \ll \gamma_c$ so that $\delta\gamma$ is {\em not} small: $\tau$ is well described by
the master curve for $p \to 0$ (see red line with circles). For $\eta = -0.0001$, $\delta \gamma$ is small 
so that for small $\alpha$ one recovers the $\alpha^{-1/4}$ behaviour of figure \ref{fig:plottot} (left) (see dark blue line with crosses). At $\alpha \to 1$
all curves collapse onto the $\eta$-independent value $1/p\sqrt{1+\gamma^2} \approx 700$.
For $\eta=-1$, $\tau$ {\em increases} with $\alpha$ until it reaches a maximum around $\alpha=1$ and then drops abruptly. (Inset, left) 
Plots of $C(t-t')$ itself, the lag-dependent correlation function, emphasize this trend with $\alpha$. 
All curves have been normalized by $C(0)$ to start at one: the amplitude $C(0)$ decreases with increasing $\alpha$ as discussed in section \ref{sec:equalcorr2}.
(Right) Relaxation time for small $p$ and $\delta\gamma$ as a function of $\alpha$, showing an interpolation between the behaviours in 
figure \ref{fig:plottot} (left) and (right). 
The red master curve with circles is expected to give a good fit for $\alpha\approx  1$ only, consistent with our results.}
\label{fig:plottb}
\end{figure}
We next discuss the behaviour of the relaxation time in the second critical region, 
$\alpha \to 1$ (i.e.\ $N^{\rm s}=N^{\rm b}$) 
and $p \to 0$ (i.e.\ $k\gg\lambda$ at fixed $\sigma_{\rm s}/\sigma_{\rm b}$). The distance from 
the critical point is therefore represented by $p$ itself and by $\delta\alpha = \alpha - 1$. As is shown in \ref{sec:pwotime2app}, when both are small
the results are independent of the degree of symmetry $\eta$ of the interactions among the hidden variables. 
For the dimensionless relaxation time we find for positive $\delta\alpha$ (see \ref{sec:pwotime2app}) 
\be
\label{eq:timep0}
\at \sim \frac{1}{\delta \alpha} \qquad p < \delta \alpha \ll 1
\ee
and this is consistent with numerical data for $\alpha>1$, see figure \ref{fig:plottot} (right). 
The figure also shows that as $\alpha$ decreases below unity the relaxation time reaches a plateau, whose height can be worked out as $1/p\sqrt{1+\gamma^2}$. 
The plateau and the decay in eq.~(\ref{eq:timep0}) are obtained as the two limiting behaviours of an $\eta$-independent master curve that applies 
when $\delta\alpha$ is of order $p$.

To understand the behaviour across the entire range $0<\alpha<1$ one can work out a separate 
master curve for fixed $\alpha$ and $p\to 0$ (see Eq. \eqref{mctimep0} in \ref{appendix:b}), which is plotted
in figure \ref{fig:plottb} (left) for $\eta=-1$ along with numerical results. 
This curve does have a nontrivial dependence both on $\eta$ and on $\alpha$; 
for $\alpha\to 0$ it approaches a constant, whose value can be checked to be consistent with the $\alpha=0$ result (Eq. \eqref{eq:pno}) as it must.

A comparison of figures \ref{fig:plottot} (right) and \ref{fig:plottb} (left) shows that the range around $\alpha=1$ 
where relaxation times are independent of $\eta$ depends not just on $p$ but also on $\gamma$. 
For small $\gamma$ as in figure \ref{fig:plottot} (right) we see good agreement with the $\eta$-independent master curves down 
to at least $\alpha\approx 0.5$. For $\gamma=1$, the value we take to plot curves for $\eta<0$, figure \ref{fig:plottb} (left) shows 
that already at $\alpha \sim 0.9$ there is visible variation across different values of $\eta$.

Figure \ref{fig:plottb} (left) contains a further insight: at fixed $\gamma$ and moderate $\alpha$, varying $\eta$ can shift the system 
away from the second critical region ($p\to 0$) towards the first one ($\delta \gamma\to 0$). Indeed, if $\gamma=1$, $\delta \gamma=\gamma_c-\gamma$ becomes
small when $\eta$ is close to zero because $\gamma_c=1/(1+\eta)$. Accordingly for the $\eta=-0.0001$ curve in the figure, 
the relaxation time for small $\alpha$ is captured better by the master curve for the first critical region (dark blue line with crosses) 
than the one for the second region. At fixed $\eta$ and for appropriate combinations of small $\delta\gamma$ and $p$ one can even see both 
critical regions as $\alpha$ is increased. Figure \ref{fig:plottb} (right) shows a case in point.

As a general trend in the relaxation times one sees that they decrease significantly with increasing $\alpha$ (see e.g.\ figure \ref{fig:plottb}, 
right): as the values of the hidden variables become constrained increasingly strongly by those of the observed ones, the 
remaining uncertainty in the prediction becomes local in time.
There are surprising counter-examples, however. Figure \ref{fig:plottb} (left) shows that for strongly anti-symmetric interactions, e.g.\ $\eta=-1$,
the relaxation time $\tau$ {\em increases} with $\alpha$ up until $\alpha=1$. This unusual trend is linked to a change of shape in the decay of 
the time-dependent correlation function, which is strongly non-exponential for $\alpha<1$ but becomes close to exponential beyond (see the inset of figure \ref{fig:plottb}, left).

\subsection{Correlation functions for $\gamma \to \gamma_c$ and $\alpha\to 0$}
\label{sec:FT1}
\begin{figure}
\includegraphics[width=0.48\textwidth]{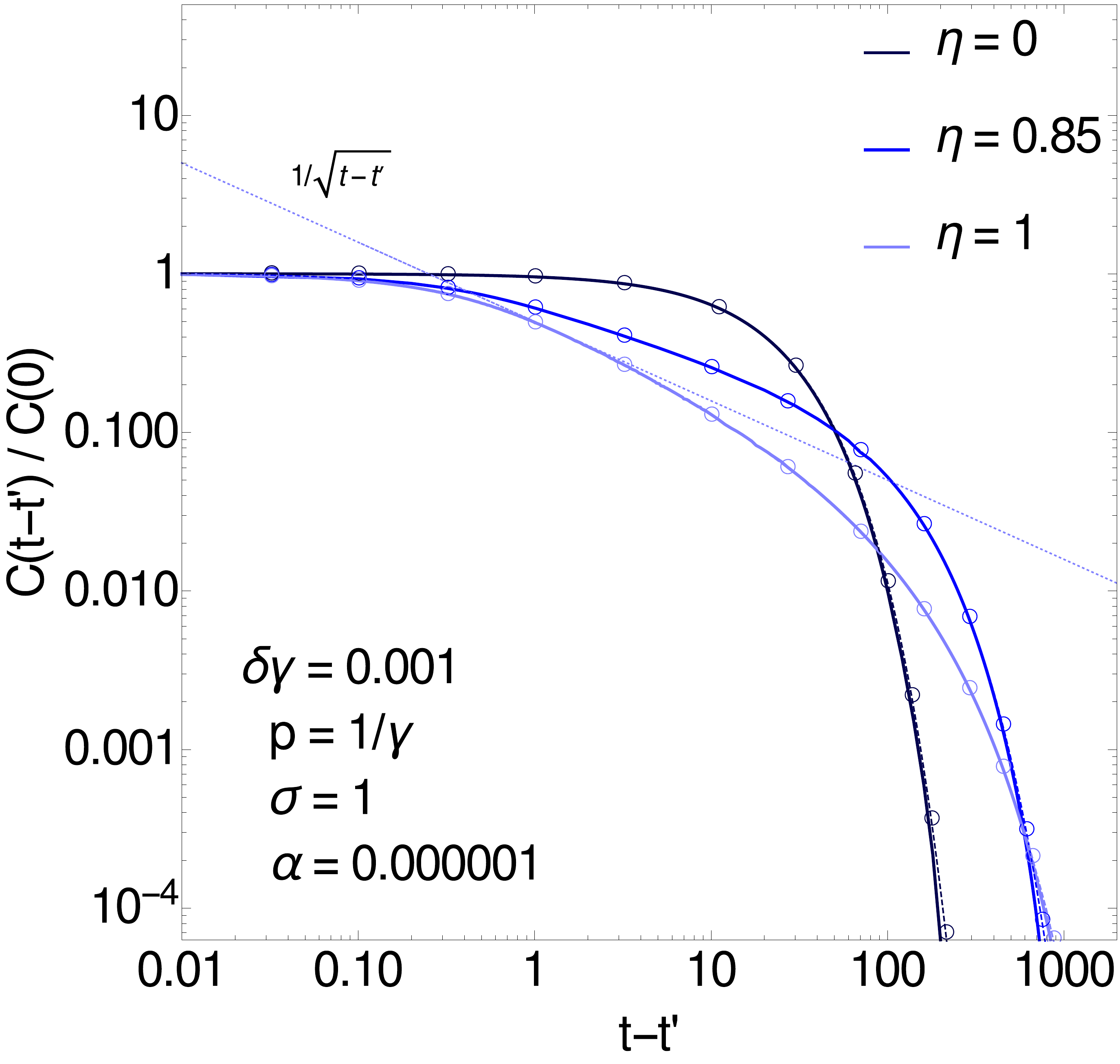}
\includegraphics[width=0.48\textwidth]{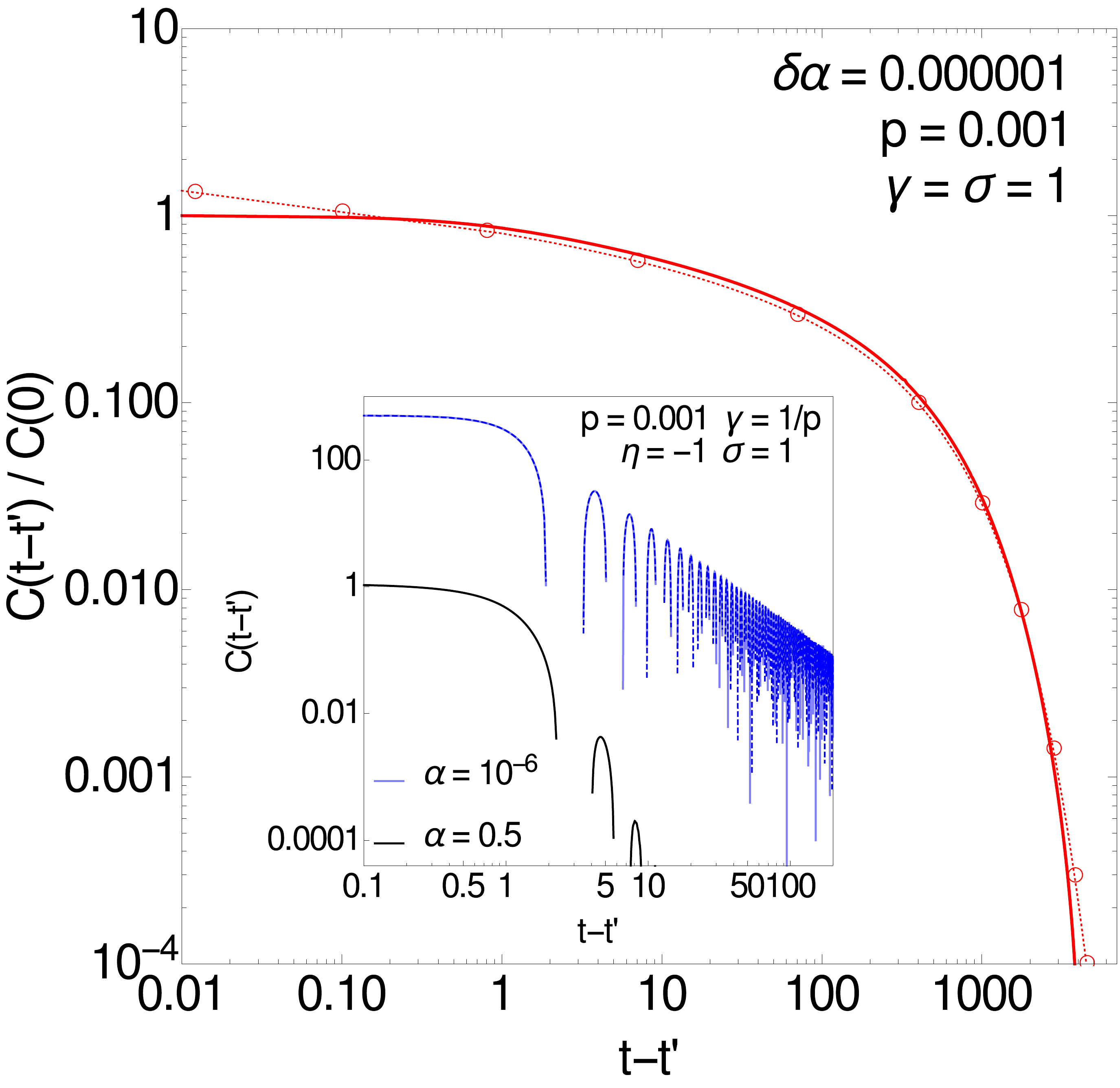}
\caption{(Left) Correlation function $C(t-t')$ in the critical region $\gamma \to \gamma_c$ and $\alpha\to 0$, for different $\eta$. 
Solid lines are the numerics, dashed ones with circles the limit curves for $\alpha=0$. Curves are normalized at the origin to focus on the shape of the 
correlation function decay. The curve for $\eta=0.85$ interpolates between the one for $\eta=1$, which has a power law regime $\sim 1/\sqrt{t-t'}$ 
(over a range that is still small in the figure but eventually diverges as $\delta\gamma\to 0$), and the pure exponential behaviour one 
finds for  $\eta=0$. The crossover occurs at $t-t'\sim 1/\epsilon^2 \sim 45$.
(Inset, right) Correlation functions for $\eta=-1$ for small $\alpha$, which are essentially identical to the  limit curve for $\alpha=0$ (dashed line), and for $\alpha=0.5$; note the oscillatory behaviour.
At $\alpha=0$ it is convenient to set $p = 1/\gamma$ (see \ref{sec:pwoMC1}) 
so we use this parameter setting also for nonzero $\alpha$ here. 
Small oscillations at long times can occur also in correlation functions at higher $\eta$, e.g.\ for $\eta=0$ in the left figure.
(Right) Correlation function in the region $p \to 0$ and $\delta \alpha \to 0$. Numerical results are essentially independent of $\eta$ so close to $\alpha=1$; we take $\eta=-0.1$. 
The master curve (dashed red line with circles) is expected to capture the behaviour for $t-t'$ beyond $1/\omr^{*} \sim 700$ but the agreement is
good also for smaller time lags. For $t-t'\to 0$ the master curve would diverge and we have therefore normalized it by the $C(0)$ of the numerical 
correlation function.}
\label{fig:FT}
\end{figure}
Having studied overall relaxation timescales, we next look more comprehensively at the time correlation functions $C(t-t')$, which can be obtained numerically by inverse Fourier transform of the power spectrum.

For the first critical region we refer to figure \ref{fig:FT} (left). Here 
the master curves are the result at $\alpha=0$ from \cite{bravi}: for $\eta=0$ and $\eta=1$ these are given 
respectively by a pure exponential and by an exponentially weighted integral of a modified Bessel function $I_1$.
In the latter case the correlation function has 
a $1/\sqrt{t-t'}$ regime indicated in the figure, before crossing over to $(t-t')^{-3/2}$ times an exponential cutoff (which kicks in around $t-t' = 250$ in the case shown in the figure).
For intermediate symmetries (e.g.\ $\eta=0.85$ in figure \ref{fig:FT} left) one has a crossover
between this behaviour and the pure exponential for $\eta=0$. This crossover occurs
at a time $\sim 1/\epsilon^2$ (here $\epsilon=1-\eta$ as before), which 
is consistent with the frequency space crossover at frequency $\sim \epsilon^2$, see e.g.\ figure \ref{fig:plotebar} (right).

For $\eta=-1$, in the $\alpha=0$ case $C(t-t')$ is a Bessel function of the first kind $J_1$ \cite{bravi}. This has oscillations, and we find that this behaviour
persists as $\alpha$ is increased.
This is shown in the inset of figure \ref{fig:FT} (right)
where the oscillations exhibit a longer decay time for $\alpha=0.5$ consistently with the findings on relaxation times of 
figure \ref{fig:plottb} (left). Note that small oscillations are also present in the curve $\alpha=0.5$ of figure \ref{fig:plottb} (left, inset) on 
longer times than shown in the graph.

\subsection{Correlation functions for $\alpha\to 1$ and $p \to 0$}
\label{sec:FT2}
We show an example of $C (t-t')$ in this second critical region in figure \ref{fig:FT} (right).  
The master curve for the power spectrum in this region (see \ref{sec:pwoMC2}) for $\delta\alpha \to 0$ tends to $ 1/\sqrt{1+(\omr/p\sqrt{\gamma^2+1})^2}$, 
independently of $\eta$. Its inverse
Fourier transform gives a modified Bessel function $K_0(\omr/p\sqrt{\gamma^2+1})$ in the time domain, which is plotted in the figure alongside the numerical results
(dashed red line with circles).

\subsection{Equal time posterior variance for $\gamma \to \gamma_c$ and $\alpha\to 0$}
\label{sec:equalcorr}
\begin{figure}
\includegraphics[width=0.47\textwidth]{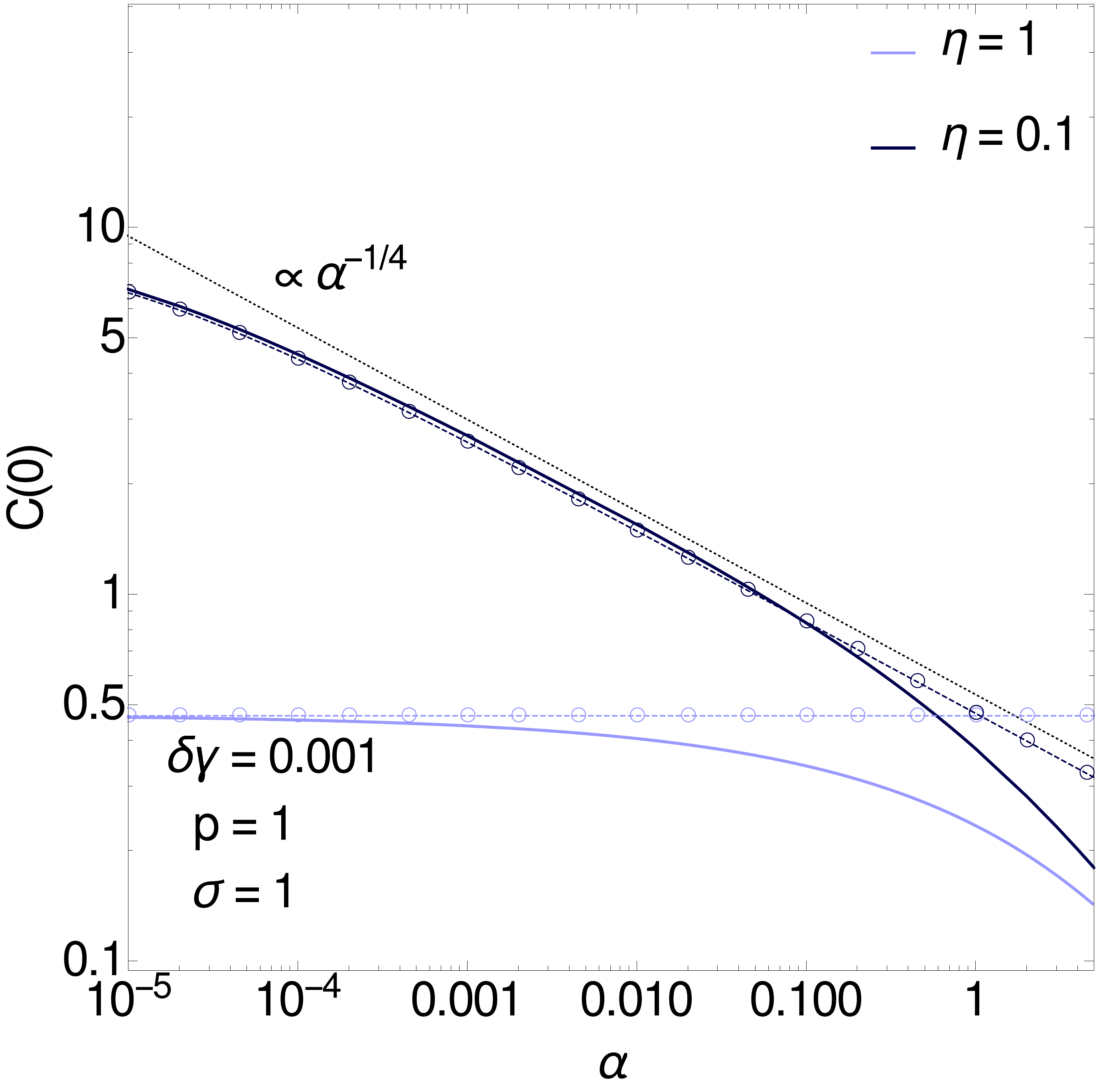}
\includegraphics[width=0.48\textwidth]{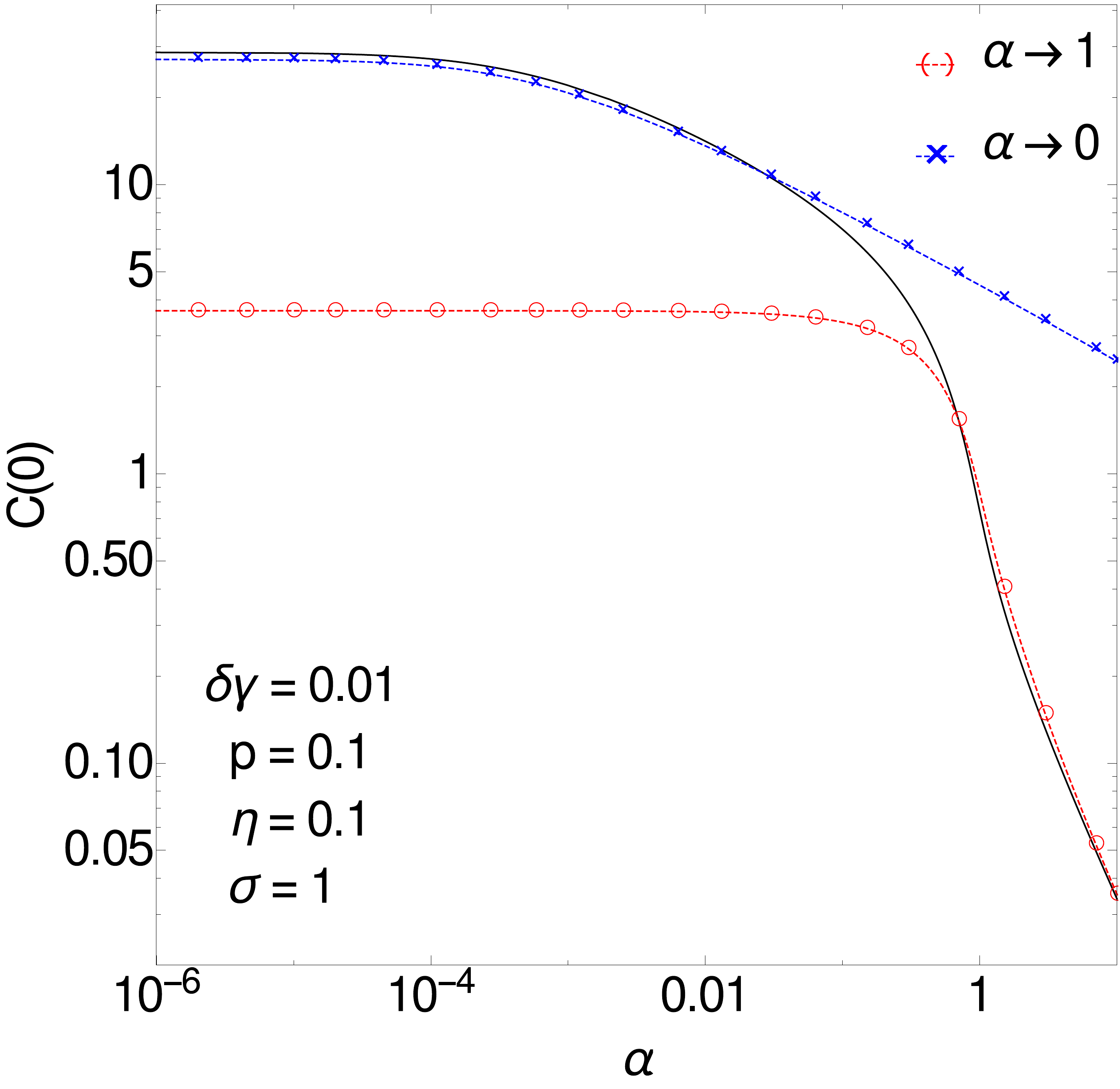}
\caption{(Left) Inference error in the vicinity of the critical region $\alpha \to 0$ and $\gamma \to \gamma_c$, as a function of $\alpha$ and for different $\eta$.
Solid lines are the numerics, dashed ones with circles the analytic master curves. The dominant contribution to the inference error is given by frequencies $\omr\sim \omr^{*}$ for $\eta=0.1$ while
it is given by $\omr\sim O(1)$ for $\eta=1$.  This leads to a power law dependence of the master curve in the former case, while in the latter regime the master ``curve'' is the $\alpha$-independent posterior variance of the case without observations, 
i.e.\ a constant, which $C (0)$ approaches for $\alpha\to 0$ as it should; for larger $\alpha$ it exhibits a smooth dependence on $\alpha$ unconnected to any critical behaviour.
(Right) Inference error for small $p$ and $\delta\gamma$ as a function of $\alpha$. This connects the behaviours in the left plot and in figure \ref{fig:integralsAmplitudep0}. The red master curve with circles is expected 
to give a good fit only around $\alpha\approx1$, as observed.}
\label{fig:integralsAmplitudea0}
\end{figure}
We turn finally to the behaviour of the inference error for the prediction of hidden unit trajectories. This is given by the equal time posterior correlator
\be
\label{eq:ampl}
C (t-t)=C (0)=\frac{1}{2\pi}\int_{-\infty}^{\infty}\tilde{C} (\omega)d\omega=\frac{1}{2\pi}\frac{\sigma_{\rm s}^2}{k^2}\sigma\int_{-\infty}^{\infty}\ac(\omr)d\,\omr=
\frac{\sigma_{\rm s}\sigma_{\rm b}}{k}\ac_{0}
\ee
where $\ac_{0}$ is a dimensionless equal time posterior variance. We see that the size of the error is generically 
proportional to the noise 
acting on the dynamics of hidden and observed variables, and inversely proportional to the hidden-to-observed interaction strength.
We summarize in the following subsections our results for $C(0)$, with 
a focus on the $\alpha$-dependence, and leave details to \ref{sec:equalcorrapp} and \ref{sec:equalcorr2app}. We begin with the first critical region.

\subsubsection{$\eta=1$.}
For $\eta=1$ we find that the equal time correlator becomes essentially independent of $\alpha$ for small $\alpha$, as one can see 
in figure \ref{fig:integralsAmplitudea0} (left, curves for $\eta=1$). More generally the result is that the 
dependence on $\alpha$ across the range $0<\alpha<1$ is smooth, and to leading order unaffected by the vicinity of the critical region.

\subsubsection{$ -1<\eta <1 $.}																										
Here one obtains (see \ref{sec:equalcorrapp}) 
\begin{eqnarray}
C(0) \sim \frac{1}{p}(1-\eta)^{\frac{7}{4}}(1+\eta)^{\frac{1}{4}} {\alpha}^{-\frac{1}{4}}
\label{eq:C0first_critical}
\end{eqnarray}
We thus predict $C(0)\sim\alpha^{-1/4}$, and this is consistent with the numerics, see e.g.\ figure \ref{fig:integralsAmplitudea0} (left, curves for $\eta=0.1$). It is notable that Eq.~(\ref{eq:C0first_critical}) is independent of $\delta\gamma$; this behaviour holds for $\delta\gamma^2\ll \alpha \ll 1$, while for smaller $\alpha$ the value of $\delta\gamma$ would become relevant.
The power law behaviour $C(0)\sim\alpha^{-1/4}$ is also consistent with the scaling $\tilde{C}(0)\sim \tau \,C (0)$ that one would expect on general grounds:
the zero frequency power spectrum $\tilde{C}(0)$ is the integral of the correlation function, hence should scale as its amplitude $C(0)$ times the 
decay time $\tau$. That this relation holds here follows from our previous result, $\tau\sim \alpha^{-1/4}$, and $\tilde{C}(0)\sim \alpha^{-1/2}$ (see \ref{sec:pwoMC1}).

\subsection{Equal time posterior variance for $\alpha\to 1$ and $p \to 0$}
\label{sec:equalcorr2}
\begin{figure}
\includegraphics[width=\textwidth]{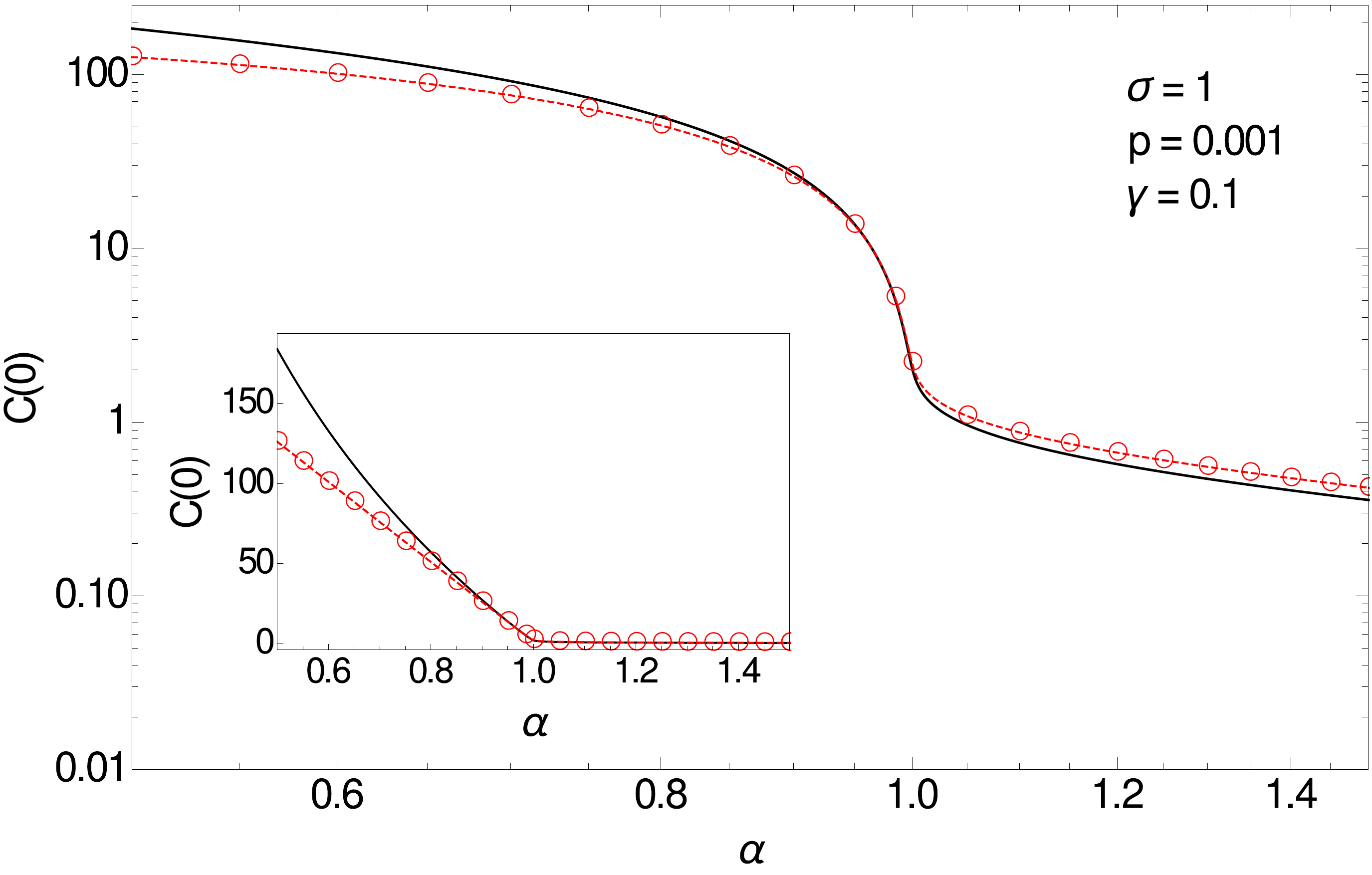}
\caption{Inference error in the vicinity of the critical region $\alpha \to 1$ and $p \to 0$, as a function of $\alpha$, at fixed $\gamma=0.1$.
The dashed line with circles shows the prediction from the logarithmic divergence and first subleading term (see equation \eqref{two_dominant}), which is qualitatively remarkably accurate even away from $\alpha=1$. The prediction
behaves as $\sim 1-\alpha$ for $\alpha<1$ as the linear scale inset shows. 
For the numerical results we used $\eta=0.1$; other $\eta$ give virtually indistinguishable curves.
}
\label{fig:integralsAmplitudep0}
\end{figure}
In the second critical region, the dominant terms of the integral \eqref{eq:ampl} can be shown to scale near $\alpha=1$ as 
$1-\alpha$ for $\alpha<1$ and $\mbox{const.}-\ln(\alpha-1)$ for $\alpha>1$ (see \ref{sec:equalcorr2app}).
These terms are plotted (red line) in figure \ref{fig:integralsAmplitudep0} where the linear scale 
inset clearly shows the linear dependence on $1-\alpha$. A separate $p\to 0$ master curve for fixed $\alpha$ below 1 can also be derived 
(by integrating the solution of \eqref{mcp0}) and matches smoothly to the $1-\alpha$ behaviour around $\alpha=1$. 

As a common trend across the two critical regions we have the intuitively reasonable result that the inference 
error decreases when the number of observed variables gets bigger. 
Figure \ref{fig:integralsAmplitudea0} (right) summarizes this and shows that, as in the case of the relaxation times, 
the behaviour with increasing $\alpha$ can cross over from the first critical region to the second provided that both $\delta\gamma$ and $p$ are small. 
It is worth re-emphasizing the non-trivial power law dependences of the inference error on $\alpha$ that occur in our dynamical setting and are quite 
distinct from the simpler behaviour in static learning scenarios~\cite{hertz,oppersolvable}.

\section{Conclusion}

In this paper we considered the problem of inferring hidden states over time in
a network of continuous degrees of freedom given a set of observed trajectories.
We started from the 
results for a linear dynamics with Langevin noise derived in \cite{plefkaobs} and in particular we
 looked at the hidden posterior variance as a measure of the inference error.
To study the average performance case, we considered the stationary regime (where time translation invariance makes it convenient to work in
Fourier space), mean-field couplings (all-to-all, weak and long-ranged) and the limit
of an infinitely large bulk size. Under these conditions, the Extended Plefka expansion used in~\cite{plefkaobs} becomes exact; also errors become 
site-independent self-consistently and equivalent 
to the \emph{average} error, which measures the quality of the prediction from the macroscopic point of view. 

Our main goal was to study the properties of this average inference error, and the associated correlation functions and timescales of the posterior dynamics, as a function of the relevant dimensionless 
parameters $\alpha,\gamma,p,\eta$. Here $\alpha$ is the ratio between observed and hidden nodes,
$\gamma$ is related to the bulk internal stability and $p$ gives the relative weight of self-interactions 
and hidden-to-observed couplings. These structural parameters are assumed to be known, either by direct measurement or 
by theoretical estimation, and our results for the posterior statistics then quantify
their interplay in determining the prediction error.

As the parameter space is relatively large, we organized the analysis around the critical regions where the (suitably non-dimensionalized) prediction error diverges.
We first studied the power spectrum of the posterior correlations, deploying critical scaling approaches to identify the 
relevant variables and find scaling functions. 
These results could then be straightforwardly translated into the corresponding ones for relaxation times and inference errors, 
providing master curves for appropriately scaled numerical data, with non-trivial power law dependences on e.g.\ $\alpha$ resulting 
from the dynamical nature of our inference scenario.
 
The first critical region we analyzed corresponds to $\alpha\ll1$, where there are many fewer observed nodes 
than hidden ones. Here we found that the presence of interaction symmetry ($\eta=1$) leads to quite different scaling behaviour than for the generic 
case $-1<\eta<1$, indicating the importance of even small deviations from detailed balance for the dynamics. 
This is in qualitative agreement with earlier studies on
systems without observations, e.g.~\cite{sompolinsky1, glassy}.

The second critical region is $0<\alpha<1$ and $p\to 0$, where some parts of the hidden dynamics are strongly constrained but because $\alpha<1$ there 
are other parts that remain unconstrained as there are still not enough observed nodes.
The resulting singularities are driven essentially by 
the hidden-to-observed couplings and are therefore 
independent of the hidden-to-hidden interaction symmetry.
A qualitative analogy can be drawn here to studies on ``underconstrained'' learning in neural networks, e.g.\
\cite{hertz,oppersolvable}, where the inference error
can diverge at the point where the number of patterns to be learned equals the number of degrees of freedom. 
This happens when no weight decay is imposed on the dynamics, which in our scenario corresponds to small $\lambda$ and hence small $p$. The analogy is only partial as our dynamical setting has a much richer range of behaviours overall.

Another interesting comparison can be made. The Extended Plefka Expansion has been applied also to spin systems for the inference of hidden 
states \cite{romanobattistin}. The analytically tractable scenario of an infinitely large network of spins with random asymmetric couplings 
was studied using a replica approach \cite{inference1} and there the error in predicting the states of hidden nodes 
does not exhibit a singularity structure like the one we find. 
It would be interesting to consider scenarios between these two extremes as 
in e.g.~\cite{sollich_BM}, to understand the relative importance of the type of dynamics (linear vs nonlinear) and 
the type of degrees of freedom (continuous vs discrete) in this context.

We stress that here the only form of noise we have included is \emph{dynamical} noise acting on the time evolution of the observed nodes, rather than measurement noise affecting the accuracy of the observed trajectory. We limited ourselves to this case to focus our analysis on the interplay of other 
parameters such as the interaction strength and the number of observations compared to the number hidden nodes. In future work it would be desirable to include measurement noise in the observation process as has been done in e.g.\ dynamical learning in neural networks \cite{krogh}.
A further extension would allow for observations that are both noisy and sparse (in time) \cite{opper_pre}.

In terms of other future work and potential applications, our results could be of interest in experimental design when 
only the spatio-temporal evolution of a few nodes can be controlled. 
Given our systematic analysis of the dependence of the average inference error on key parameters of the system,
one could study how this might guide the experimental set-up in such a way as to maximize the inference accuracy.
For example, the parameter $\alpha$ measuring the relative number of nodes to monitor over time
could enter the specification of a hypothetical experimental protocol. We recall that, for both critical regions, we found that
the inference error decreases with this parameter and identified the relevant power laws.
If we suppose that an estimate of other parameters ($\gamma$, $p$, $\eta$) is available either from previous measurements or some a priori knowledge, 
then our explicit expressions for $C(0)$, the average inference error when many hidden trajectories are reconstructed, 
might serve to fix a minimal $\alpha$ needed for achieving a $C(0)$ below a set precision threshold. 

A major, complementary problem when extracting information from data is 
the estimation of parameters, as well as identifiability \cite{stumpf_robustness}
and ultimately model selection. Statistical physics-inspired techniques
have already been successfully applied e.g.\ to signalling and regulatory networks 
\cite{perturbations, braunstein} for learning the couplings from steady state data.
To see how our results could be relevant in this regard, note that tackling inverse 
problems relies on an interplay between state inference and parameter estimation.
In this paper, we have analyzed the inference problem for the time courses of hidden nodes, 
assuming that the model parameters are randomly distributed with known average properties, such as
the coupling strength and the degree of symmetry (i.e.\ the deviation from equilibrium).
As a next step one could consider inferring the parameters by Expectation-Maximization \cite{em},
where the Expectation part relies precisely on computing the posterior statistics. 
Our simple expressions for the average posterior variance in terms of the average coupling strength and degree of symmetry
would then simplify this procedure in some regimes and help investigate it in an analytically controlled, thus more insightful, way.

\section*{Acknowledgement}
This work was supported by the Marie Curie Training Network NETADIS (FP7, grant 290038). 
The authors acknowledge the stimulating research environment provided by the EPSRC Centre for Doctoral Training in Cross-Disciplinary Approaches 
to Non-Equilibrium Systems (CANES, EP/L015854/1). BB acknowledges also the Simons Foundation Grant No. 454953.

\cleardoublepage

\appendix
\section{Scaling analysis: power spectra}
\label{appendix:a}

\subsection{Master curves for $\gamma\to\gamma_c$ and $\alpha\to 0$}
\label{sec:pwoMC1}
\begin{figure}
\includegraphics[width=0.475\textwidth]{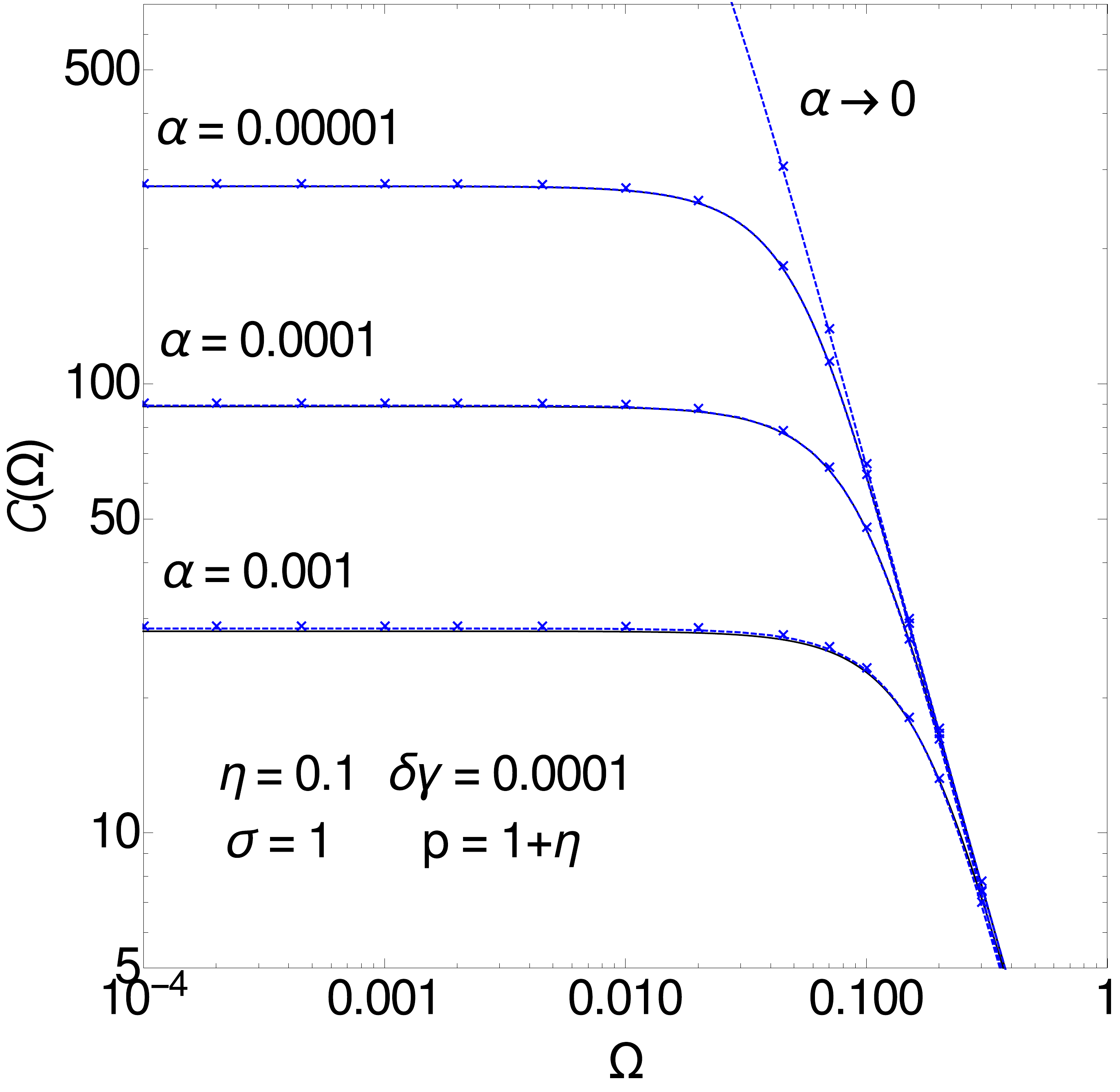}
\includegraphics[width=0.48\textwidth]{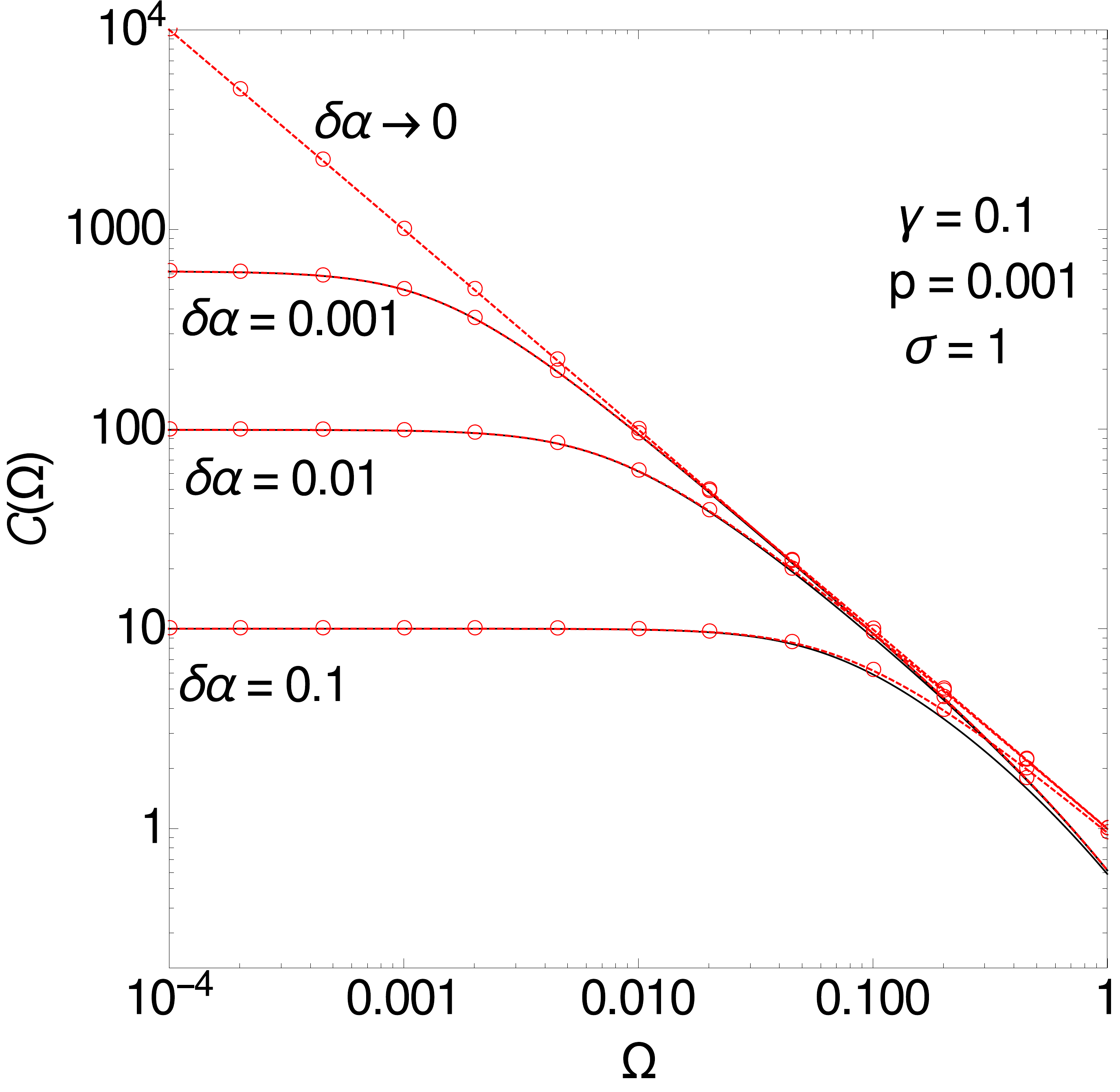}
\caption{(Left) Numerically calculated power spectra for small $\alpha$, showing the approach to the limit curve for $\alpha\to 0$ with $\gamma$ close to $\gamma_c$ (first critical region). 
This master curve and the ones 
for $\alpha=0.00001$, $\alpha=0.0001$ and $\alpha=0.001$ are given by blue dashed lines with crosses. In this way one can see the amplitude variation for relatively large values of $\bar{\alpha}$; in the plot  $\bar{\alpha}\sim10^2$ (for $\alpha=0.00001$), 
$10^3$ (for $\alpha=0.0001$) and $10^4$ (for $\alpha=0.001$). Even for smaller $\bar{\alpha}$, the variation in shape with this parameter is small: $\bar\alpha$ mainly affects the height of the plateau close 
to $\omr=0$ and the position of the crossover to the large frequency Lorentzian tail.
(Right) Power spectra for small $\delta\alpha=\alpha-1$, illustrating the approach to the limit curve for $\delta\alpha\to 0$ with $p$ close to $0$ (second critical region). This master curve and the ones 
for $\delta\alpha=0.001$, $\delta\alpha=0.01$ and $\delta\alpha=0.1$ are shown as red 
dashed lines with circles. From this plot one can examine the variation 
with $\delta\bar\alpha$, which in the plot has 
values $\delta\bar\alpha\sim1$ (for $\delta\alpha=0.001$), $\delta\bar\alpha\sim10$ (for $\delta\alpha=0.01$) and $\delta\bar\alpha\sim100$ (for $\delta\alpha=0.1$). 
For the numerics we used $\eta=0.5$, though again the curves are essentially $\eta$-independent.}
\label{fig:alpha0}
\end{figure}
We begin with the behaviour of the posterior covariance power spectrum for $\gamma\to\gamma_c$ and $\alpha\to 0$.
We already know the limit curve for the power spectrum: it is given by the spectrum {\em at} the critical point. 
At this point we have an $\alpha=0$ system, i.e.\ without observations \cite{bravi}. As the interactions and noise level relating to observed nodes are then irrelevant, we can set 
$k=j$, $\sigma_{\rm b}=\sigma_{\rm s}$ (i.e $p=1/\gamma$, $\sigma=j$ and $\Omega=\omega/j$).
From the expression in \cite{bravi}, the dimensionless power spectrum then reads
\begin{equation}
\label{eq:pno}
\ac_{\gamma,\eta}(\omr)=\frac{4}{\bigg[\frac{1}{\gamma}+\text{i}\omr+\sqrt{\big(\frac{1}{\gamma}+\text{i}\omr\big)^2-4\eta}\bigg]\bigg[\frac{1}{\gamma}-\text{i}\omr+\sqrt{\big(\frac{1}{\gamma}-\text{i}\omr\big)^2-4\eta}\bigg]-\frac{4}{\sigma^2}}
\end{equation}
$\ac(0)$ becomes singular when $\gamma=\gamma_c=1/(1+\eta)$, as \eqref{eq:pno} then has a pole at $\omr=0$. The approach to 
$\eqref{eq:pno}$ at decreasing $\alpha$ is plotted in figure \ref{fig:alpha0} (left).

\subsubsection{$\eta=1$.}
\begin{figure}
\includegraphics[width=0.487\textwidth]{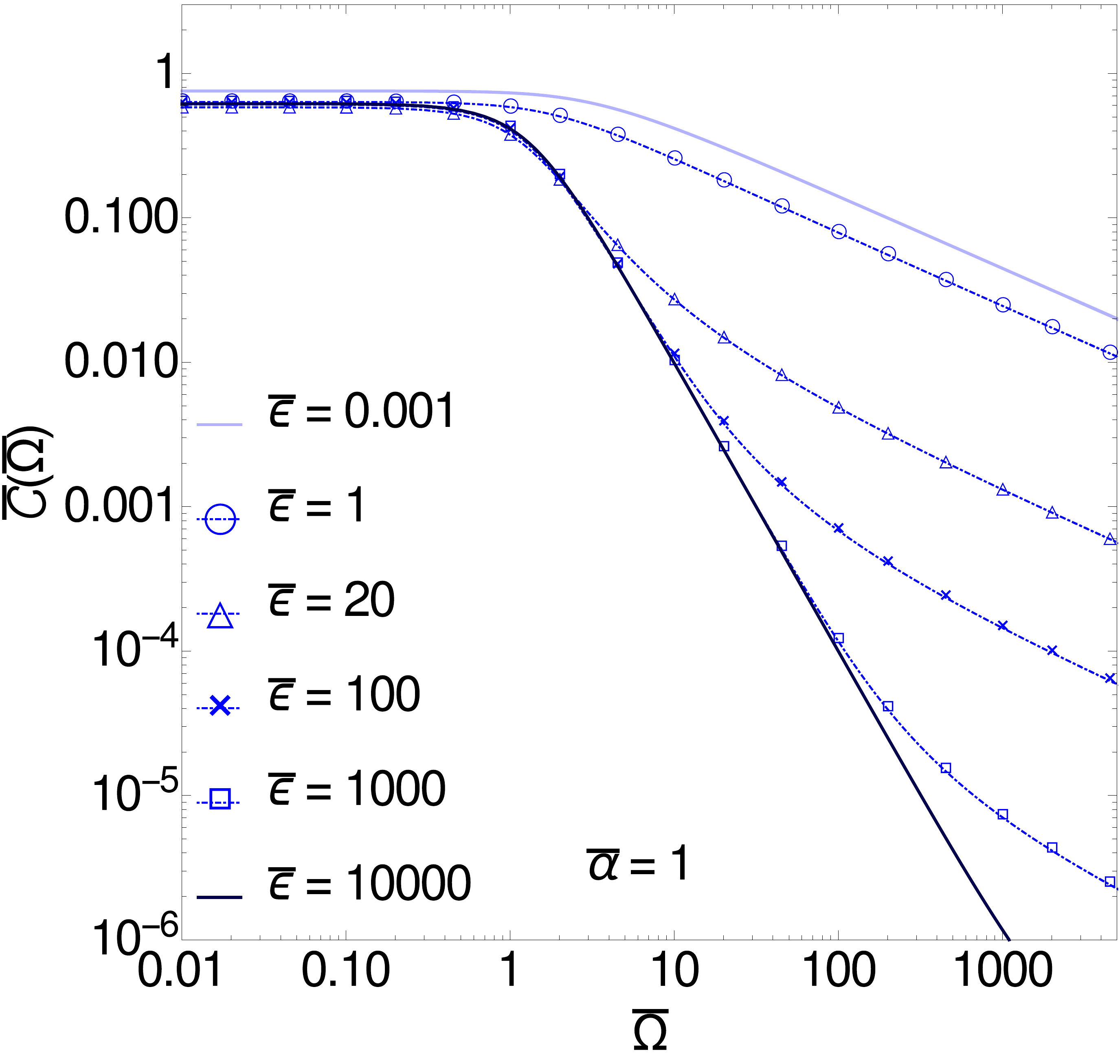}
\includegraphics[width=0.48\textwidth]{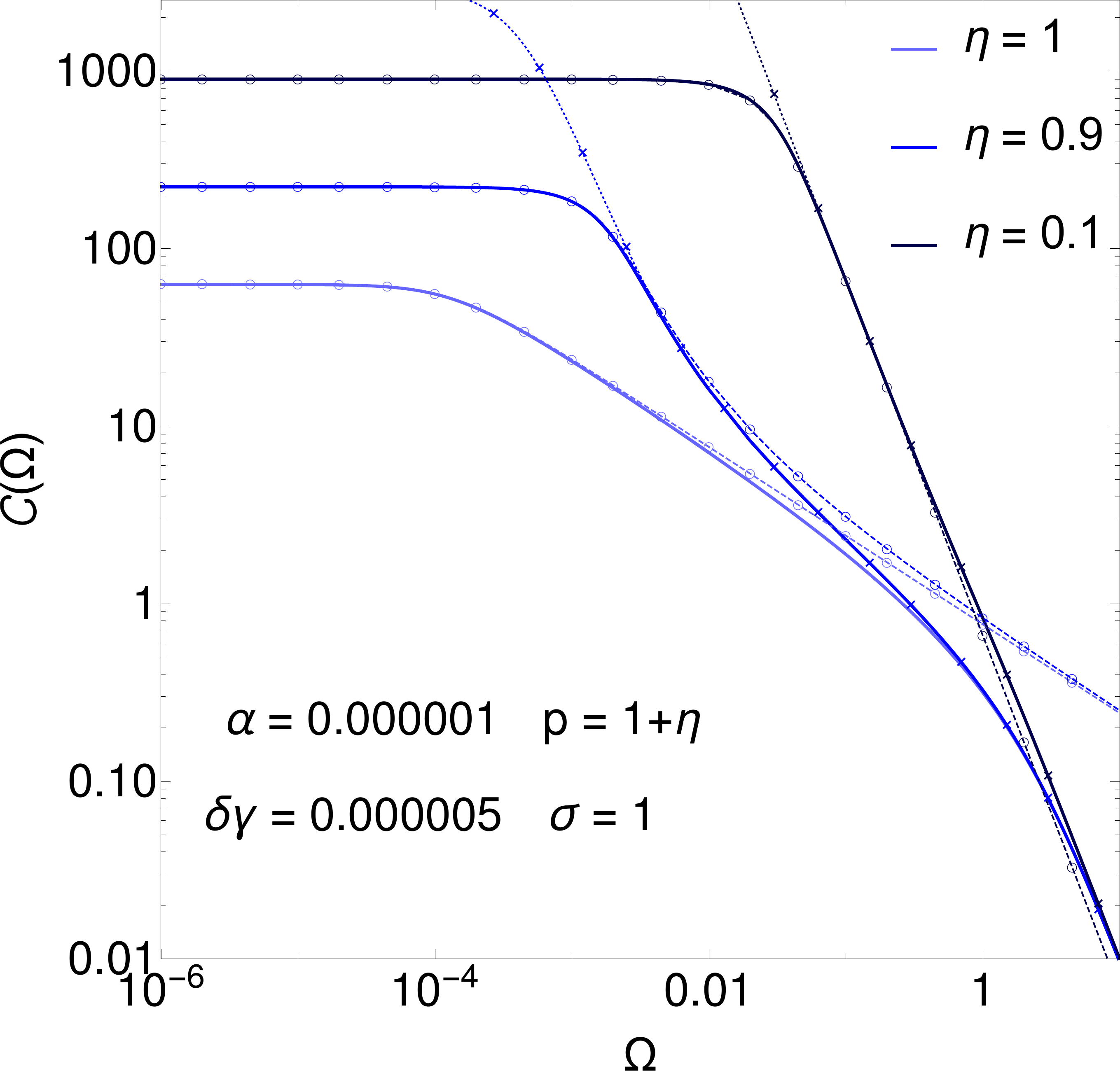}
\caption{(Left) Master curves for different values of $\bar{\epsilon}$, the parameter indicating effectively the distance from $\eta=1$ when $\delta\gamma$ is close to zero.
One can see the $1/\sqrt{\bar{\omr}}$ tail for $\bar{\epsilon}\to 0$ and the $1/\bar{\omr}^2$ one for $\bar{\epsilon}\to \infty$ (light blue and dark blue curves), while intermediate values
of $\bar{\epsilon}$ show a crossover between these two tails.
(Right) Numerically determined power spectra for different values of $\eta$, the symmetry parameter, at $\alpha \to 0$ and $\delta\gamma \to 0$. For
$\eta=0.1$, $\omr^{*}\sim 3\cdot10^{-3}$ and for $\omr^{*}\ll \omr \ll 1$ one sees the Lorentzian tail. For $\eta=1$, $\omr^{*}\sim 5\cdot 10^{-6}$ and in the range $\omr^{*}\ll \omr \ll 1$ one has the $\sim 1/\sqrt{\omr}$ tail.
For $\eta=0.9$ (i.e.\ $\epsilon=1-\eta=0.1$, $\bar\epsilon \approx 45$), the results interpolate between these two regimes as expected from the left figure; the crossover occurs at $\omr\sim 0.01$.
In fact, the limit $\bar{\ac}|_{\bar{\omr}\to \infty}\sim 1/\bar{\omr}^2$, once one reinserts the dependence on $1-\eta = \epsilon$, gives $\ac(\omr) \sim \epsilon^3/\omr^2$ for $\omr^{*} \ll \omr \ll 1$, while \eqref{eq:mc1} has a tail 
$\ac(\omr) \sim 2/\sqrt{\omr}$: these two tails meet around $\omr\sim \epsilon^2$ ($\epsilon^2=0.01$ in this case), in agreement with the results for the limit $\epsilon \ll 1$ in the case 
without observations \cite{bravi}. 
The dashed lines with circles are master curves for $\omr$ in the vicinity of $\omr^{*}$ for the different $\eta$. 
Dotted lines with crosses trace the master curves at $\omr \sim O(1)$, which are  independent of $\alpha$ and equal to the curves at $\alpha=0$. In the relevant range of large frequencies they 
behave essentially as Lorentzians.}
\label{fig:plotebar}
\end{figure}

To explain the rescaling procedure for understanding the approach to the singularity at $\gamma=\gamma_c$ when $\alpha$ is nonzero, we first focus on the case $\eta=1$.
In section \ref{sec:eta1} we derived an algebraic equation for $\tilde{C}(z)$ which, for the power spectrum $\mathcal{C}(\Omega)$, becomes
\begin{equation}
\label{eq:adimsymps}
 -\omr^2=-\frac{1}{\ac}+\frac{\alpha}{1+\ac}+\frac{(\gamma p)^2}{1+(\gamma p)^2 \ac}+\frac{p^2}{\big[1+2(\gamma p)^2 \ac\big]^2}
\end{equation}
Approaching the singularity along either of the two directions $\alpha\to 0$ or $\delta\gamma\to 0$ we get two distinct power law 
divergences of $\ac(0)$; a third direction of approach is from nonzero $\Omega$ at $\alpha=\delta\gamma=0$.
Approaching along the $\delta\gamma$-direction ($\delta \gamma \to 0$ at $\alpha=\Omega=0$) we find from \eqref{eq:adimsymps}
\begin{equation}
\label{eq:scalgc}
\ac(0)\sim\frac{1}{p^2\sqrt{\delta\gamma}} 
\end{equation}
For $\alpha\to 0$ at $\Omega=0$ and $\gamma=\gamma_c$, with $\gamma_c=1/2$ as $\eta=1$, the result is
\begin{equation}
\label{eq:scala}
 \ac(0)\sim \frac{{\alpha}^{-\frac{1}{3}}}{p^2}
\end{equation}
Finally for $\Omega\to 0$ at $\alpha=\delta\gamma=0$ the low frequency tail of $\ac(\omr)$ is known from \cite{bravi} as
\begin{equation}
\label{eq:tailsym}
 \ac(\omr)\sim\frac{1}{\sqrt{\omr}}
\end{equation}
These three power laws can be combined
into the scaling 
\begin{eqnarray}
 \ac(\omr)&=&\frac{1}{p^2\sqrt{\delta\gamma}}\,\bar{\ac}(\bar{\omr},\bar{\alpha})\label{rescsymm1}\\
 \bar{\omr}&=&\frac{\omr }{2\, p\,\delta \gamma}\label{rescsymm2}\\
 \bar{\alpha}&=&\frac{\alpha}{16\,\delta\gamma^{\frac{3}{2}}}\label{rescsymm3}
\end{eqnarray}
where the exponents of $\delta\gamma$ in $\bar\Omega$ and $\bar\alpha$ are fixed by standard arguments
(see \cite{thesis} for details). Inserting this ansatz 
into \eqref{eq:adimsymps} and taking the limit $\delta\gamma\to 0$ gives
\begin{equation}
\label{eq:mc1}
 -1 + \bar{\ac}^2+ \bar{\alpha}\,\bar{\ac}^3 + \frac{\bar{\omr}^2}{4}\,\bar{\ac}^4=0
\end{equation}
This equation implicitly determines the master curve, i.e.\ the scaling function $\bar{\ac}(\bar{\omr},\bar{\alpha})$.
It describes the power spectrum in the region where $\delta\gamma$, $\alpha$ and $\Omega$ are all small but $\bar\Omega$ and $\bar\alpha$ are finite, which requires in
particular that the dimensionless frequency $\Omega$ must be of the order of $\delta \gamma$. 
Numerical data in figure \ref{fig:plotebar} (right) show good agreement with this master curve in the relevant regime.

\subsubsection{$-1<\eta<1$.}
Let us now consider $\eta<1$, for which it is convenient to work with the entire system \eqref{eq:adimplefka}. For $\delta\gamma \to 0$ at $\alpha=0$ one finds amplitudes 
$\ac(0) \sim (1-\eta)/2\,\delta\gamma \,p^2$ and $\abb(0)\sim -4 \,\delta\gamma^2 (1+\eta)^4/p^4$.
At $\delta\gamma=\Omega=0$, we get $\ac(0) \sim \alpha^{-1/2}$. 
For the third direction we can read off from \cite{bravi} that at $\alpha=\delta\gamma=0$, the low-frequency tail of the power spectrum is 
given by $1/\omr^2$ for $\eta < 1$. 
Comparing the first and third expression suggests a crossover frequency determined by $1/\omr^{2}=(1-\eta)/2\,\delta\gamma \,p^2$, giving
$\omr \sim p\sqrt{2\,\delta \gamma}$. Using this we define scaling functions for $\mathcal{C}$ and $\mathcal{B}$ as
\begin{eqnarray}
 \ac(\omr)&=&\frac{1-\eta}{2\,\delta\gamma \,p^2}\bar{\ac}(\bar{\omr},\bar{\alpha})\label{resc01}\\
\abb(\omr)&=&-\frac{4\, \delta\gamma^2 (1+\eta)^4}{1-\eta^2}\bar{\abb}(\bar{\omr},\bar{\alpha})\label{resc02}\\
 \bar{\omr}&=&\frac{\omr}{\omr^{*}}=\frac{\omr}{p\,(1-\eta)}\sqrt{\frac{1+\eta}{2\,\delta\gamma}}\label{resc03}\\
 \bar{\alpha}&=&\frac{\alpha\,(1-\eta)}{4\,\delta\gamma^2(1+\eta)^3}\label{resc04}
\end{eqnarray}
The somewhat complicated looking prefactors are chosen here to give scaling functions that will be independent of $p$ and
$\eta$. The response $\mathcal{R}$, which also features in the original equations
\eqref{eq:adimplefka}, does not need to be rescaled as it turns out to be equal to unity to leading order.

We insert the above rescalings into the system \eqref{eq:adimplefka} and again look at the limit $\delta\gamma\to 0$. Some care is needed as there are competing orders of $\delta\gamma$ in the equations 
so that one has to expand the response $\mathcal{R}$ not just to $\mathcal{O}(1)$ but to $\mathcal{O}(\delta\gamma)$.
One then finds simply $\bar{\abb}(\bar{\omr},\bar{\alpha})=\bar\alpha$ and this makes sense: at $\alpha=0$ we must retrieve 
the results of \cite{bravi}, where the normalization of the MSRJD path integral leads all moments of auxiliary
variables to vanish. The master curve for the posterior covariance spectrum can also be obtained explicitly, as
\begin{equation}
\label{eq:mc2}
\bar{\ac}(\bar{\omr},\bar{\alpha})=\frac{1}{2\bar{\alpha}}\bigg[-(1+\bar{\omr}^2)+\sqrt{4\bar{\alpha}+(1+\bar{\omr}^2)^2}\bigg]
\end{equation}
It has the limits $\bar{\ac}|_{\bar{\alpha}\to 0}\sim 1/(1+\bar{\omr}^2)$, $\bar{\ac}|_{\bar{\omr}\to \infty}\sim 1/\bar{\omr}^2$ and 
$\bar{\ac}|_{\bar{\alpha}\to \infty}\sim 1/\sqrt{\bar{\alpha}}$. The latter tells us how the prediction error decreases in the regime where $\alpha$ is still small but larger 
than $\delta \gamma^2$. 
The fit provided by the master curve \eqref{eq:mc2} for different small values of $\alpha$ is shown in figure \ref{fig:alpha0} (left).

\subsubsection{Crossover at $\eta\approx 1$.}
Above we found different power law behaviours and scaling functions for $\eta=1$ and $\eta< 1$, thus a 
crossover must occur at $\eta\approx 1$. To see this, the two cases $\eta=1$ and $\eta< 1$ can be analyzed as limit cases of a more general scaling ansatz that accounts explicitly for
the effect of $\epsilon=1-\eta$, i.e.\ the distance from the symmetric value $\eta=1$. This becomes an additional critical parameter that enters the scaling functions in the combination
\be
\bar{\epsilon} = \epsilon/\sqrt{\delta\gamma}
\ee
We define these scaling functions via
\begin{eqnarray}
 \ac(\omr)&=&\frac{(2+\bar{\epsilon})}{2\, \sqrt{\delta\gamma}\,p^2}\bar{\ac}(\bar{\omr},\bar{\alpha},\bar{\epsilon})\\
\abb(\omr)&=&-\frac{32\,\delta\gamma^{\frac{3}{2}}}{2+\bar{\epsilon}}\bar{\abb}(\bar{\omr},\bar{\alpha},\bar{\epsilon})\\
 \bar{\omr}&=&\frac{\omr}{p\,(2+\bar{\epsilon})\delta\gamma}\\
 \bar{\alpha}&=&\frac{\alpha\,(2+\bar{\epsilon})}{32\,\delta\gamma^{\frac{3}{2}}}
\end{eqnarray}
We then find again $\bar{\mathcal{B}}=\bar\alpha$ while
the master curve $\bar{\ac}(\bar{\omr},\bar{\alpha},\bar{\epsilon})$ solves a fourth-order equation
\be
\label{mcebar}
(8+\bar{\ac}\,\bar{\epsilon}(2+\bar{\epsilon}))^2(-4+\bar{\ac}(2+\bar{\epsilon})(-\bar{\epsilon}+\bar{\ac}(1+\bar{\alpha}\,\bar{\ac})(2+\bar{\epsilon})))+\bar{\ac}^4(2+\bar{\epsilon})^6\bar{\omr}^2=0
\ee
The solution of this for a range of different $\bar{\epsilon}$ is plotted in figure \ref{fig:plotebar} (left).

The two previous cases $\eta=1$ and $\eta< 1$ (with $\delta\gamma\to 0$) are recovered as the limits respectively for $\bar{\epsilon}\to 0$ and $\bar{\epsilon}\to\infty$. 
In the first limit, $(2+\bar{\epsilon})\sqrt{\delta\gamma}\sim 2\sqrt{\delta\gamma}$ and the rescaling relations \eqref{rescsymm1}, \eqref{rescsymm2}, \eqref{rescsymm3} 
for $\eta=1$ are retrieved as they should be; accordingly, the equation \eqref{mcebar} for the master curve becomes exactly \eqref{eq:mc1}. On the other hand, when $\bar{\epsilon}\to \infty$, $(2+\bar{\epsilon})\sqrt{\delta\gamma}\sim \bar{\epsilon}\sqrt{\delta\gamma}=\epsilon$ and we recover the
rescalings $\eqref{resc01}$-$\eqref{resc04}$ adopted for $\eta< 1$; in this case \eqref{mcebar} reduces to
\be
-1+\bar{\ac}+\bar{\alpha}\bar{\ac}^2+\bar{\ac}\bar{\omr}^2=0
\ee
whose positive solution is given by \eqref{eq:mc2}.

\subsection{Master curves for $\alpha \to 1$ and $p \to 0$}
\label{sec:pwoMC2}
In this section we look at the scaling around the second critical region, $\alpha \to 1$ and $p \to 0$. 
At $\alpha=1$, we find the following power law scaling with $p$ of the amplitudes
\begin{eqnarray}
&&\ac(0) \sim \frac{1}{p\sqrt{\gamma^2+1}}\\
&&\ar(0)\sim \frac{\gamma p}{\sqrt{\gamma^2+1}}\\
&&\abb(0)\sim -1
\end{eqnarray}
Approaching from the other direction, $p=0$, gives $\ac(0)\sim 1/\delta\alpha$. At $\alpha=1$ and $p=0$, finally, one finds for small nonzero 
frequency $\Omega$ that $\ac(\Omega)\sim 1/\Omega$.
Equating the three divergences above identifies a crossover frequency $\omr^{*}=p\sqrt{\gamma^2+1}$ and similarly a characteristic value for $\delta\alpha$. 
We thus define rescaled quantities again
\begin{eqnarray}
\ac(\omr)&=&\frac{1}{p\,\sqrt{\gamma^2+1}}\,\bar{\ac}(\bar{\omr},\delta\bar\alpha)\label{rescp1}\\ 
\ar(\omr)&= &\frac{\gamma\, p}{\sqrt{\gamma^2+1}}\,\bar{\ar}(\bar{\omr},\delta\bar\alpha)\label{rescp2}\\
\mathcal{\hc}(\omr)&=&-\bar{\mathcal{\hc}}(\bar{\omr},\delta\bar\alpha)\\
\bar{\omr}&=&\frac{\omr}{\omr^*}=\frac{\omr}{p\,\sqrt{\gamma^2+1}}\label{rescp3}\\
\delta\bar\alpha&=&\frac{\delta \alpha}{p\,\sqrt{\gamma^2+1}}\label{rescp4}
\end{eqnarray}
Inserting into the system \eqref{eq:adimplefka}, taking $p\to 0$ and keeping only the leading terms one finds as the master curve for $\mathcal{C}(\Omega)$
\begin{equation}
\label{eq:mc3}
\bar{\ac}(\bar{\omr},\delta\bar\alpha)=\frac{-\delta\bar\alpha+\sqrt{4+\delta\bar\alpha^2+4\bar{\omr}^2}}{2\,(1+\bar{\omr}^2)}
\end{equation}
with limits $\bar{\ac}|_{\delta\bar\alpha\to 0}\sim 1/\sqrt{1+\bar{\omr}^2}$, $\bar{\ac}|_{\bar{\omr}\to \infty} \sim 1/\bar{\omr}$ and
$\bar{\ac}|_{\delta\bar\alpha\to \infty}\sim 1/\delta\bar\alpha$. From the latter one sees again the decrease of the inference error for increasing number of observations 
(while remaining in the regime studied here where $\delta\alpha$ is small, $p<\delta\alpha\ll1$ ). The master curve predictions for generic $\delta\bar\alpha$ are compared to direct numerical evaluation 
of $\mathcal{C}$ in figure \ref{fig:alpha0} (right).

So far we had focussed on the $\alpha\to 1$ end of the second critical region.
As this region covers the entire range $0<\alpha<1$, however, one can also study the critical behaviour as $p\to 0$ for fixed $\alpha<1$. The crossover into this region 
can be seen by taking $\delta\bar\alpha\to -\infty$, 
where $\bar{\ac}\to |\delta\bar\alpha|/(1+\bar{\omr}^2)$ from \eqref{eq:mc3}. Including the prefactor from \eqref{rescp1} and using \eqref{rescp3} gives 
\begin{equation}
\label{mc1m}
\ac(\Omega)=\frac{1-\alpha}{p^2(\gamma^2+1)+\Omega^2} 
\end{equation}
This suggests that in general, for finite $1-\alpha$, $\ac$ will be $\sim 1/p^2$. Generalizing this suggests the following scaling for small $p$ and
$0<\alpha<1$ 
\begin{eqnarray}
\ac(\omr)&=&\frac{1}{(p\,\gamma)^2}\,\bar{\ac}(\bar{\omr},\delta\alpha,\eta)\label{rescp12}\\ 
\ar(\omr)&= &\gamma\,\bar{\ar}(\bar{\omr},\delta\alpha,\eta)\label{rescp22}\\
\mathcal{\hc}(\omr)&=&-\bar{\mathcal{\hc}}(\bar{\omr},\delta\alpha,\eta)\\
\bar{\omr}&=&\frac{\omr}{p}\label{rescp32}
\end{eqnarray}
with $\delta\alpha=1-\alpha$.
By substitution into the system \eqref{eq:adimplefka} and taking the limit $p\to 0$ one can then obtain explicit 
solutions for $\bar{\ac}$, $\bar{\ar}$, $\bar{\mathcal{\hc}}$. In particular we find $\bar{\mathcal{\hc}}=\alpha$, similarly to the previous
cases, while $\bar{\ac}$ satisfies
\be
\label{mcp0}
\frac{\alpha-1}{\bar{\ac}}+\frac{1}{1+\bar{\ac}}+\frac{1}{\gamma^2}\bigg[\frac{1}{(1+\bar{\ac}(1+\eta))^2}+\frac{\bar{\omr}^2}{(1+\bar{\ac}(1-\eta))^2}\bigg]
\ee
In the limit where $1-\alpha\ll 1$ one retrieves the result \eqref{mc1m} as required for consistency between the two scaling limits.

The result above is of interest for negative symmetry parameters $\eta$. As one approaches the extreme case $\eta \to -1$, i.e.\ close to antisymmetry, the critical $\lambda$ at $\alpha = 0$ tends to zero. As a consequence, 
for finite $k$, one has that the approach to the stability limit of the hidden dynamics corresponds to $p\to 0$ and is therefore captured by the master curve \eqref{mcp0}.

\cleardoublepage
\section{Scaling analysis: timescales and amplitudes}
\label{appendix:b}

\subsection{Relaxation time for $\gamma\to\gamma_c$ and $\alpha\to 0$}
\label{sec:pwotimea1app}
The relaxation time is defined in a mean-squared sense by \eqref{eq:time} and can be rewritten in terms of the rescaled power spectrum as 
\be
\tau^2=-\frac{1}{2\, \bar{\ac}(0,\bar{\alpha})}
\frac{d^2 \bar{\ac}(\bar{\omr},\bar{\alpha})}{d^2\omega}\bigg\lvert_{\omega=0}
\ee
We consider the dimensionless version of this typical timescale \eqref{eq:timeadim}, $\at = \sigma\tau$,
and analyze separately the vicinity of the two critical points. As $\tau$ is determined directly from the power spectrum, its scaling behaviour follows from that of
$\mathcal{C}(\Omega)$. 
Explicitly, in terms of the dimensionless frequency $\Omega =\omega/\sigma$, which for critical scaling is rescaled further to $\bar{\omr}=\omr/\omr^*$, one has
\be
\label{eq:timeapp}
\at = \at^{*}\bar{\tau}(\bar{\alpha}), \qquad
\bar{\tau}^2(\bar{\alpha}) = -\frac{1}{2\, \bar{\ac}(0,\bar{\alpha})}
\frac{d^2 \bar{\ac}(\bar{\omr},\bar{\alpha})}{d^2\bar{\omr}}\bigg\lvert_{\bar{\omr}=0}
\ee
with $\at^{*}=1/\omr^{*}$.

\subsubsection{$\eta=1$.}
The relaxation time is rescaled using \eqref{eq:timeapp} as 
\be
\label{eq:times}
\at=\frac{1}{p\,\delta\gamma}\, \bar{\tau}(\bar{\alpha})
\ee
where $\bar{\tau}(\bar{\alpha}) $ is the solution of a system of two equations
\be
\label{timesym}
\begin{cases}
&\big(2 + 3\,\bar{\alpha}\,\bar{\ac}\big)\,\bar{\ac}\,\bar{\tau}^2-\frac{1}{4}\bar{\ac}^3=0\\
&-1 +\,\bar{\ac}^2+\bar{\alpha}\,\bar{\ac}^3=0
\end{cases}
\ee
which can be obtained by deriving from \eqref{eq:mc1} one equation for $\bar{\ac}(0,\bar{\alpha})$ and one 
for $d^2 \bar{\ac}(\bar{\omr},\bar{\alpha})/d^2{\bar{\omr}}\big\lvert_{\bar{\omr}=0}$ and using relations
\eqref{eq:timeapp} and \eqref{eq:times}. For simplicity we have denoted $\bar{\ac}(0,\bar{\alpha})$ and $\bar{\tau}(\bar{\alpha})$ as $\bar{\ac}$ and $\bar{\tau}$. From
these two equations we note $\bar{\ac}(0,\bar{\alpha})|_{\bar{\alpha}\to \infty}\sim \bar{\alpha}^{-\frac{1}{3}}$ and
$\bar{\tau}|_{\bar{\alpha}\to \infty}\sim \bar{\alpha}^{-\frac{2}{3}}$, which implies for $\delta\gamma^2 < \alpha\ll 1$ the power law \eqref{timealpha0mt}.

\subsubsection{$-1<\eta<1$.}
By applying \eqref{eq:timeapp} with \eqref{eq:mc2}, one can rescale in this regime according to
\be
\at=\frac{1}{(1-\eta)\,p}\sqrt{\frac{1+\eta}{2\, \delta \gamma}}\, \bar{\tau}(\bar{\alpha})
\ee
and $\bar{\tau}(\bar{\alpha})$ is given by
\be
\label{eq:tauh0}
\bar{\tau}(\bar{\alpha}) =\frac{1}{(4\bar{\alpha}+1)^{\frac{1}{4}}}
\ee
with the limit $\bar{\tau}|_{\bar{\alpha}\to \infty}\sim \bar{\alpha}^{-\frac{1}{4}}$ corresponding to \eqref{timealpha1mt}.

\subsubsection{Crossover at $\eta\approx 1$.}
\label{sec:crossover_time}
The relaxation time scalings above can be seen as limit cases of a more general scaling linked to the parameter $\bar{\epsilon}$. From the general master curve \eqref{mcebar} we can
derive the following system of two equations
\begin{eqnarray}
\fl 
\begin{cases}
\label{mcebar0}
 &[8+\bar{\ac}\bar{\epsilon}(2+\bar{\epsilon})]^2[-4+\bar{\ac}(2+\bar{\epsilon})(-\bar{\epsilon}+\bar{\ac}(1+\bar{\alpha}\,\bar{\ac})(2+\bar{\epsilon}))]=0\\
 &[-16\bar{\epsilon}+\bar{\ac}(2+\bar{\epsilon})(16-3\bar{\epsilon}^2+4\,\bar{\ac}\bar{\epsilon}(2+\bar{\epsilon})+
\bar{\alpha}\,\bar{\ac}(24+5\,\bar{\ac}\bar{\epsilon}(2+\bar{\epsilon}))]\bar{\tau}^2=\frac{\bar{\ac}^3(2+\bar{\epsilon})\bar{\epsilon}^4}{(8+\bar{\ac}\bar{\epsilon}(2+\bar{\epsilon}))}
\end{cases}
\end{eqnarray}
where $\bar{\ac}$ is shorthand for $\bar{\ac}(0,\bar{\alpha},\bar{\epsilon})$ and $\bar{\tau}$ for $\bar{\tau}(\bar{\alpha},\bar{\epsilon})$.
Consistent with the role of $\bar{\epsilon}$ discussed above, one can check that the limit for $\bar{\epsilon}\to 0$ is precisely the system \eqref{timesym}, 
while for $\bar{\epsilon}\to \infty$ \eqref{mcebar0} becomes
\be
\begin{cases}
\label{mcebarasy}
& -1 +\bar{\ac}+\bar{\alpha}\,\bar{\ac}^2=0\\
&-\bar{\ac}+\bar{\tau}^2(-3+\bar{\ac}(4+5\,\bar{\alpha}\,\bar{\ac}))=0
\end{cases}
\ee
Here the first equation agrees with \eqref{eq:mc2} as it should. Solving for $\bar{\tau}$, one finds \eqref{eq:tauh0}.

\begin{figure}
\includegraphics[width=0.9\textwidth]{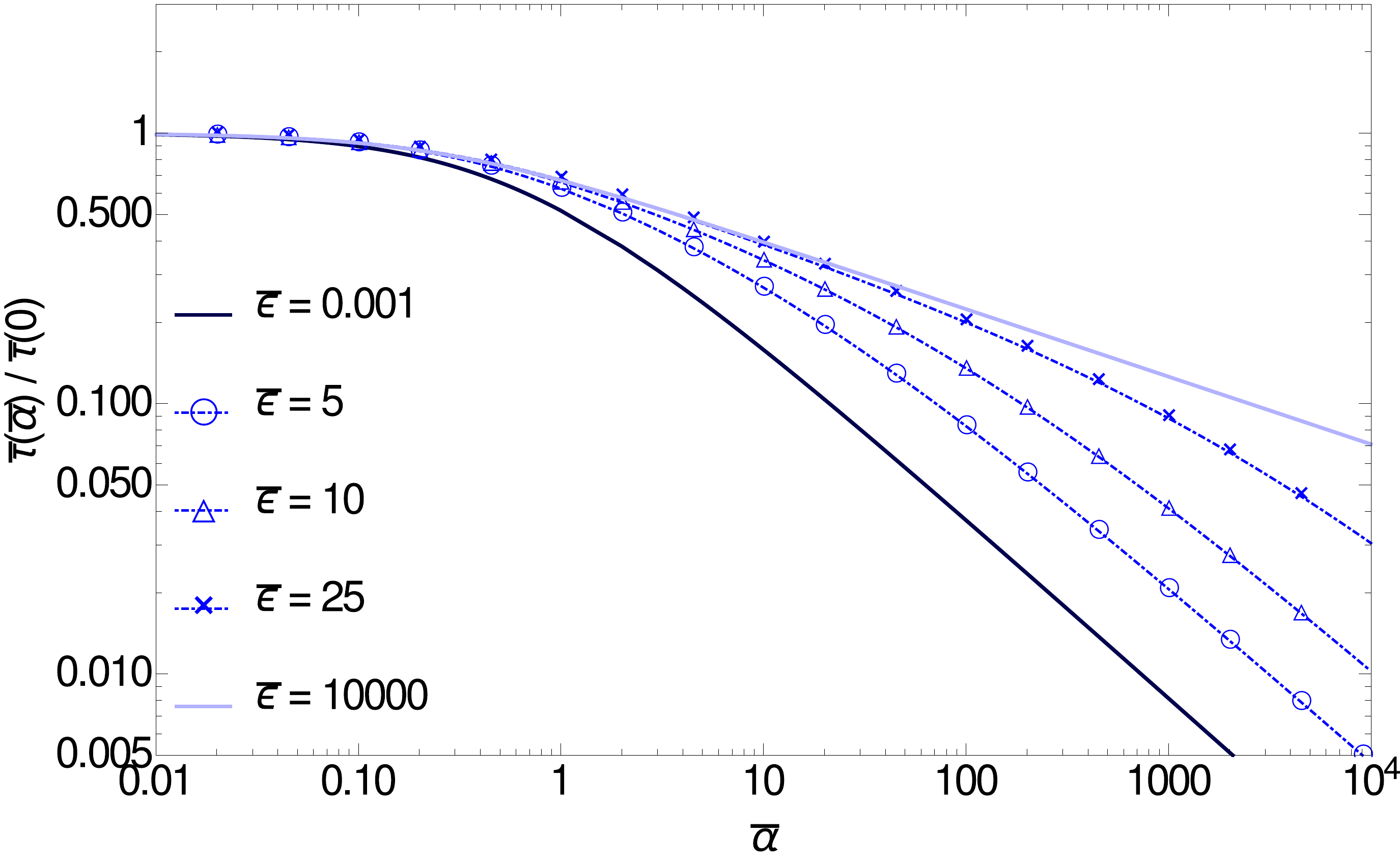}
\caption{The rescaled relaxation time $\bar{\tau}(\bar{\alpha})$, normalized at the origin, for $\alpha \to 0$ and $\gamma \to \gamma_c$ as a 
function of $\bar{\alpha}$ for different values of $\bar{\epsilon}$. The two limits,
$\bar{\epsilon}\to \infty$ and $\bar{\epsilon}\to 0$, are highlighted in light blue and dark blue respectively. 
For intermediate values of $\bar{\epsilon}$ one can see an interpolation between the corresponding
power laws, i.e.\ $\bar{\alpha}^{-\frac{1}{4}}$ and $\bar{\alpha}^{-\frac{2}{3}}$, and 
this crossover occurs at $\bar{\alpha}^{*} = \bar{\epsilon}^2/4$ (for instance, at $\bar{\epsilon}=25$, 
$\bar{\alpha}^{*} \sim 150$). In the main text, the curve for $\eta=0.85$ shown in figure \ref{fig:plottot} (left)
has a crossover at $\alpha^{*} \sim 0.0005$, which would correspond to the crossover for $\bar{\epsilon}\sim 20$ here.}
\label{fig:plottotapp}
\end{figure}

For generic $\bar{\epsilon}$, the rescaled relaxation time $\bar{\tau}(\bar{\alpha})$ must then exhibit a crossover in its large $\bar{\alpha}$ power law behaviour, i.e.\ 
from $\bar{\alpha}^{-\frac{1}{4}}$ to $\bar{\alpha}^{-\frac{2}{3}}$. This crossover takes place around $\bar{\alpha}^{*} = \bar{\epsilon}^2/4$; see figure 
\ref{fig:plottotapp}.

\subsection{Relaxation time for $\alpha \to 1$ and $p \to 0$}
\label{sec:pwotime2app}
We next consider the behaviour of the relaxation time in the second critical region. From the rescaled expression of $\mathcal{C}$ \eqref{eq:mc3} one has
\be
\label{eq:tp0}
\at=\frac{1}{p\,\sqrt{(1+\gamma^2)}}\,\bar{\tau}(\delta\bar\alpha)
\ee
and
\be
\bar{\tau}(\delta\bar\alpha)=\sqrt{\frac{1}{2}\bigg(1-\frac{\delta\bar\alpha}{\sqrt{4+\delta\bar\alpha^2}}\bigg)}
\ee
with the limit $\bar\tau|_{\delta\bar{\alpha}\to \infty}\sim 1/\delta\bar\alpha$ giving \eqref{eq:timep0}.

In the region of fixed $0<\alpha<1$ and $p\to 0$ one again has a separate master curve, where one defines a rescaled time as
\be
\at = \frac{1}{p}\bar\tau(\alpha,\eta)
\ee
using \eqref{rescp32}. Given the master curve \eqref{mcp0}, $\bar\tau$ can be found as the solution of the system 
\begin{gather}
 \label{mctimep0}
\begin{cases}
&1/\gamma^2(1+\bar{\ac}(1-\eta))^2 + (\alpha-1)\bar\tau^2/\bar{\ac} + \bar{\ac}\bar\tau^2/(1+\bar{\ac})^2+
2\,\bar{\ac}(1+\eta)\bar\tau^2/\gamma^2(1+\bar{\ac}(1+\eta))^3=0\\
&(\alpha-1)/\bar{\ac}+1/(1+\bar{\ac})+1/\gamma^2(1+\bar{\ac}(1+\eta))^2=0
\end{cases}
\raisetag{1\baselineskip}
\end{gather}

As expected, for $\alpha \to 1^{-}$ one recovers the $\delta\bar{\alpha}\to -\infty$ limit of 
\eqref{eq:tp0}, i.e.\ $\at \sim 1/p\sqrt{1+\gamma^2}$, for any $\eta$, as can be seen also 
in figure \ref{fig:plottb} (left) of the main text.

\subsection{Equal time posterior variance for $\gamma \to \gamma_c$ and $\alpha\to 0$}
\label{sec:equalcorrapp}
Finally we look at the equal time posterior correlator given by \eqref{eq:ampl} in the main text. Its critical scaling properties depend on whether the integral over $\Omega$ that defines $\ac_0$ is dominated by small frequencies $\Omega\sim\Omega^*$, where $\Omega^*$ is the relevant frequency scale 
in the appropriate critical region, or by $\Omega\sim 1$. One has the first case when the critical master curve for 
$\mathcal{C}(\Omega)$ has an integrable tail towards large (scaled) frequencies, otherwise the second.
This illustrates how the analysis in the frequency domain is important in determining the power scaling of $C(0)$.

\subsubsection{$\eta=1$.}	
Here the dominant contribution to the integral \eqref{eq:ampl} comes from $\omr\sim O(1)$, and in this region the master curve for 
the spectrum is given by equation \eqref{eq:pno} (the case without observations). This explains the effective independence from $\alpha$ pointed out
in the main text.

\subsubsection{$ -1<\eta <1 $.}	
Here the dominant contribution to the prediction error comes from $\omr\sim\omr^{*}$, thus one has to evaluate the integral of the master curve \eqref{eq:mc2}.
To do so we note from \eqref{resc01} that $\ac_{0}$ can be written in scaled form as
\be
\ac_{0} = \frac{(1-\eta)\omr^{*}}{2\,\delta\gamma \,p^2}\,\bar{\ac}_{0}(\bar{\alpha})
\label{C0_rescaled}
\ee
with $\bar{\ac}_{0}(\bar{\alpha})=\int_{-\infty}^{\infty}\bar{\ac}(\bar{\omr},\bar{\alpha})d\,\bar{\omr}$. This function encodes the entire $\alpha$-dependence in the critical region. It has a finite limit for $\bar\alpha\to 0$ while for large $\bar \alpha$ it decays as $\bar{\ac}_{0}\sim \bar\alpha^{-1/4}$ 
as one can show by noting that the relevant frequencies $\bar\omr$ in \eqref{eq:mc2} are then $\sim \bar\alpha^{-1/4}$. Using this 
asymptotic behaviour in \eqref{C0_rescaled} and substituting also the expressions for $\omr^{*}$ and $\bar\alpha$ from \eqref{resc03} and \eqref{resc04}, respectively, 
one obtains
\begin{eqnarray}
C(0) \sim \frac{(1-\eta)^2}{p\sqrt{2 \delta\gamma (1+\eta)}}\bar{\alpha}^{-\frac{1}{4}}= \frac{1}{p}(1-\eta)^{\frac{7}{4}}(1+\eta)^{\frac{1}{4}} {\alpha}^{-\frac{1}{4}}
\end{eqnarray}
i.e.\ equation \eqref{eq:C0first_critical} in the main text. Thus one can derive $C(0)\sim\alpha^{-1/4}$, where $\bar{\alpha}\gg 1$ corresponds
to $\delta\gamma^2\ll \alpha \ll 1$ from \eqref{resc04}.

\subsection{Equal time posterior variance for $\alpha\to 1$ and $p \to 0$}
\label{sec:equalcorr2app}
In the second critical region and focussing on $\alpha\to 1$, we have an interesting marginal case where the equal-time variance
\eqref{eq:ampl} has contributions from all frequencies ranging from the critical frequency scale $\omr\sim \omr^*$ to $\omr\sim 1$. 
This is because the power spectrum \eqref{eq:mc3} for critical frequencies has a $1/\bar\omr$ tail for large $\bar\omr = \omr/\omr^*$, which gives a logarithmically divergent integral \eqref{eq:ampl}. This divergence is cut off only by the crossover to a Lorentzian tail when $\omr=\mathcal{O}(1)$. Including the prefactor from \eqref{rescp1}, one thus estimates
\be
\label{dom2}
\ac_0\approx \frac{\omr^{*}}{p\sqrt{\gamma^2+1}}
\,2\int_{0}^{1/\omr^*} \bar{\ac}(\bar{\omr},\delta\bar\alpha) d\bar\omr
\ee
The fraction in front of the integral equals unity as $\omr^{*}=p\sqrt{\gamma^2+1}$ from \eqref{rescp3} so from the $1/\bar{\omr}$ tail one finds $\ac_0 \approx 2 \ln(1/\omr^*)$ to leading order. All of the interesting dependence on $\alpha$ is in the next subleading term, which is relevant in practice as it only competes with a logarithmic divergence. Writing $\ln(1/\omr^*)$ as $\int_0^{1/\omr^*} d\bar\omr/(1+\bar\omr)$, this subleading term can be split off in the form
\be
\ac_0\approx 
\,2\ln(1/\omr^*) + 2\int_0^{\infty} [\bar{\ac}(\bar{\omr},\delta\bar\alpha)-(1+\bar\omr)^{-1}] d\bar\omr
\label{two_dominant}
\ee
The remaining integral is convergent at the upper limit so we have taken the upper limit $1/\omr^*$ to infinity as is appropriate to get the 
leading contribution for $p\to 0$. The integral is then a function of $\delta\bar\alpha$ only, which one finds varies 
as $|\delta\bar\alpha|$ for $\delta\bar\alpha\to-\infty$ and as $\mbox{const.}-\ln(\delta\bar\alpha)$ for $\delta\bar\alpha\to \infty$.

Finally for the small $\alpha$ end of this critical region, one should take the integral of the solution of \eqref{mcp0}. Here the only dependence on $\alpha$ arises 
from \eqref{mcp0} itself, and gives a smooth variation of $\bar{\ac}$ with this parameter. Thus, independently of whether the integral determining the posterior variance $C(0)$ is 
dominated by $O(1)$ frequencies ($\eta=1$) or by frequencies in the vicinity of $\omr^{*}$ ($\eta<1$), the result will also be a smooth function of $\alpha$.

\cleardoublepage
\section*{References}
\bibliography{./Phdbib}

\end{document}